\documentclass[aps,pre,showkeys,longbibliography,groupedaddress,amsmath,amssymb,floatfix,twocolumn]{revtex4-1}
\usepackage{color} 
\usepackage{times}
\usepackage{todonotes}
\usepackage{verbatim}
\usepackage{bm}

\begin{document}

%----------------------------- HEADER ----------------------------------
\title{Discordant synchronization patterns on directed networks of identical phase oscillators with attractive and repulsive couplings}
\author{Thomas Peron}
\email{thomaskaue@gmail.com}
\affiliation{Instituto de Ci\^encias Matem\'aticas e de Computa\c{c}\~ao, Universidade de S\~ao Paulo,
S\~ao Carlos 13566-590, S\~ao Paulo, Brazil}

%-----------------------------------------------------------------------

%-----------------------------------------------------------------------
\begin{abstract}
We study the collective dynamics of identical phase oscillators on globally coupled networks whose interactions are asymmetric and mediated by positive and negative couplings. We split the set of oscillators into two interconnected subpopulations. In this setup, oscillators belonging to the same group interact via symmetric couplings while the interaction between subpopulations occurs in an asymmetric fashion. By employing the dimensional reduction scheme of the Ott-Antonsen (OA) theory, we verify the existence of traveling-wave and $\pi$-states, in addition to the classical fully synchronized and incoherent states. Bistability between all collective states is reported. Analytical results are generally in excellent agreement with simulations; for some parameters and initial conditions, however, we numerically detect chimera-like states which are not captured by the OA theory.    
\end{abstract}

%\pacs{05.40.-a, 05.45.Xt, 87.10.Ca}
\maketitle 

\section{Introduction}

The Kuramoto model of coupled phase oscillators has become over the years a paradigmatic tool for the study of emergent synchronization
phenomena in nonlinear sciences. In its first formulation, 
globally coupled oscillators interact via the sine of the differences of their phases; this interaction is weighted by a positive coupling strength, and by increasing its magnitude, the phases are gradually pulled towards a common value, creating then a synchronization phase transition~\cite{acebron2005kuramoto,rodrigues2016kuramoto,arenas2008synchronization}. Initially conceived as a solvable extension of the model proposed by Winfree~\cite{winfree1967biological}, the Kuramoto model attracted great attention thanks to its analytical tractability and its later discovered potential to describe synchronization phenomena in a diverse set of systems, such as in optomechanical cells~\cite{heinrich2011collective}, Josephson junctions~\cite{wiesenfeld1996synchronization}, 
chemical oscillators~\cite{kiss2002emerging}, power networks~\cite{dorfler2014synchronization}, and even the synchrony among violin players~\cite{shahal2020synchronization}.  For a long list of examples of the use of Kuramoto models in real applications see the reviews in Refs.~\cite{acebron2005kuramoto,rodrigues2016kuramoto,arenas2008synchronization}.

Many variations of the original model by Kuramoto have been inspired by particular features found in different physical systems~\cite{acebron2005kuramoto}. One example is the seminal work carried out by Daido~\cite{daido1992quasientrainment}, who, inspired by spin-glass models, treated the couplings between oscillators in a Kuramoto model as random variables which could be either positive or negative. Daido's results provided evidence for an analogous glass phase transition in oscillatory systems; however, the precise conditions for the existence of those ``oscillator glasses'' have remained unclear, and still some debate surrounds the problem~\cite{bonilla1993glassy,stiller1998dynamics,daido2000algebraic,stiller2000self,iatsenko2014glassy}. After the early works by Daido and others~\cite{daido1992quasientrainment,bonilla1993glassy,stiller1998dynamics,daido2000algebraic,stiller2000self,zanette2005synchronization}, the discussion on oscillator glasses was brought back to attention by Hong and Strogatz, who in a series of papers~\cite{hong2011kuramoto,hong2011conformists,hong2012mean} further exploited the role of negative and positive couplings. In the first coupling setting considered by them~\cite{hong2011kuramoto,hong2011conformists}, oscillators were divided into two globally
coupled subpopulations characterized by distinct coupling strengths. In a scenario resembling sociodynamical models, oscillators within the first subpopulation were modeled to have the tendency to align with the mean-field (conformist oscillators), whereas the second subpopulation was defined by oscillators that are repelled by the
other units (contrarian oscillators). In the second model~\cite{hong2012mean}, a fraction of the oscillators was considered to provide positive coupling inputs to other nodes, while the remaining contributed with negative couplings; that is, in mathematical terms, the coupling variable was placed inside the summation term of the interaction function. 
%a scheme akin to inhibitory and excitatory interactions in neuronal systems. 
Despite being very similar, the two coupling formulations have been shown to yield significantly different collective dynamics; in fact, only 
the model in Refs.~\cite{hong2011kuramoto,hong2011conformists} was found to lead to 
different transitions other than between incoherence and 
classical partially synchronized states. 

Although Hong and Strogatz did not bring new evidence to support or discard  the existence of oscillator-glasses, their papers motivated several other studies on discordant synchronization patterns -- i.e., states characterized by the separation of the population of oscillators into partially synchronized clusters --  induced by the coexistence of attractive and repulsive couplings (see, e.g., ~\cite{iatsenko2014glassy,montbrio2011collective,iatsenko2013stationary,hong2014periodic,kloumann2014phase,sonnenschein2015collective,ottino2018volcano,park2018metastable,anderson2012multiscale,park2020competing}). Of particular interest here is the work by Sonnenschein et al.~\cite{sonnenschein2015collective}, where the authors unified the coupling settings of Refs.~\cite{hong2011kuramoto,hong2011conformists,hong2012mean} into a single model that also included the influence
of stochastic fluctuations on the frequencies. More specifically, in Ref.~\cite{sonnenschein2015collective}, Kuramoto oscillators were set to interact concomitantly via  a coupling $K_i$, which was placed outside the summation term of the interaction function, regulating the neighboring influence perceived by oscillator $i$; and a coupling $G_i$ placed inside the sum, endowing the oscillators with the ability to contribute differently to the mean-field. By employing the dimension reduction framework offered by the Gaussian Approximation~\cite{sonnenschein2013approximate}, the authors showed that all states
previously reported in Refs.~\cite{hong2011kuramoto,hong2011conformists,hong2012mean} (namely, traveling waves and $\pi$-states, and conventional incoherent and synchronous states) persisted in the model with both 
types of couplings, but under new routes outlining the transitions between different states.

The relevance of mixing positive and negative couplings in phase-oscillator models actually goes beyond the theoretical interest in oscillator glasses: It turns out that certain types of physical, biological and chemical systems can indeed be described as oscillators coupled through attractive and repulsive interactions. Noteworthy real-world examples showing similar characteristics and phenomena to those described above include laser arrays~\cite{tradonsky2015conversion,pal2020rapid} and electrochemical oscillators~\cite{kori2018partial,sebek2019plasticity}.  Furthermore, the balance between phase-attraction and -repulsion has been recently shown to be a key factor in the regulation of circadian rhythms by pacemakers cells in the suprachiasmatic nucleus (SCN)~\cite{myung2015gaba}. 

Motivated by the aforementioned contributions, here we investigate the model in Ref.~\cite{sonnenschein2015collective} of identical oscillators in the absence
of stochastic fluctuations acting on the frequencies. We divide the oscillators into two subpopulations asymmetrically coupled, and employ the theory by Ott and Antonsen (OA)~\cite{ott2008low,pikovsky2015dynamics} to obtain a reduced set of equations that describes the evolution of the system. By studying the linear stability of the reduced system, we analytically derive several conditions that delineate the transitions between synchronized, incoherent, traveling-waves and $\pi$-states. Interestingly, we find the dynamics of the present model to be overall more intricate than its stochastic version~\cite{sonnenschein2015collective}, with wider regions in the parameter space exhibiting coexistence between different synchronization patterns. As we shall see, simulations with large populations of oscillators in general confirm with excellent agreement the predictions by the theory; for a small set of parameters, however, we report strong deviations from the dynamics yielded by the reduced system.

\section{Model}

Following Sonnenschein et al.~\cite{sonnenschein2015collective}, we study here the system made up of $N$ identical Kuramoto oscillators whose equations are given by
\begin{equation}
\dot{\theta}_{i}=\omega_{0}+\frac{K_{i}}{N}\sum_{j=1}^{N}G_{j}\sin(\theta_{j}-\theta_{i}),
\label{eq:KGmodel}
\end{equation}
where $i=1,...,N$, and $\omega_0$ is the natural frequency. Notice that, in contrast to Ref.~\cite{sonnenschein2015collective}, we do not consider identical oscillators under the influence of stochastic fluctuations in the phase dynamics; instead, the only source of disorder is the one inflicted by the coupling strengths. We henceforth refer to parameters $K_i$ and $G_j$ as the in- and out-coupling strengths, respectively. As defined in Eq.~(\ref{eq:KGmodel}), these couplings set the interactions between the oscillators to be asymmetric (or directed): oscillator $i$ contributes to the dynamics of neighboring nodes with weight $G_i$, while the input 
arising from other oscillators is weighted by $K_i$. The first model by Hong and Strogatz~\cite{hong2011kuramoto} is recovered when $G_i=1 \forall i$ in Eq.~(\ref{eq:KGmodel}) (no out-coupling strengths), while the second model investigated by the same authors~\cite{hong2012mean} is obtained by symmetrizing the in-coupling strengths, i.e., $K_i = 1 \forall i$. Bifurcation conditions have been calculated recently for a stochastic system  with a coupling scheme similar to Eq.~(\ref{eq:KGmodel})~\cite{meylahn2020two}. Other similar forms of the coupling setting of Eq.~(\ref{eq:KGmodel}) have also been addressed recently in Refs.~\cite{vlasov2014synchronization,iatsenko2014glassy}, considering a phase frustration term in the interaction function (Kuramoto-Sakaguchi model~\cite{sakaguchi1986soluble}), and in populations of asymmetrically coupled R\"ossler oscillators~\cite{peron2016traveling}.  

\section{Dimensional reduction}
\label{sec:dim_reduction}
In the continuum limit $N \rightarrow \infty$, we rewrite the original Eq.~(\ref{eq:KGmodel}) by omitting the sub-indexes as
\begin{equation}
\dot{\theta}=\omega_0+KR\sin(\Theta-\theta), 
\end{equation}
where $R$ and $\Theta$ are the ``weighted'' order parameter and the mean-field phase, respectively, defined by
\begin{equation}
Re^{i\Theta}=\int\int Gr_{K,G}e^{i\psi_{K,G}}P(K,G)dKdG. 
\label{eq:total_order_parameter}
\end{equation}
$P(K,G)$ is the joint distribution of in- and out-coupling strengths; variables $r_{K,G}e^{i\phi_{K,G}}$ are the local order parameters that quantify the synchrony within subpopulations: 
\begin{equation}
z_{K,G} = r_{K,G}e^{i\psi_{K,G}}=\int\rho(\theta,t|K,G)e^{i\theta} d\theta,
\label{eq:rKG}
\end{equation}
where $\rho(\theta,t|K,G)$ is the probability density function  of observing an oscillator with phase $\theta$ at time $t$ for a given coupling pair $(K,G)$. Henceforth we adopt the notation $\rho_{K,G}(\theta,t) \equiv \rho(\theta,t|K,G)$. The normalization condition $\int_{-\pi}^{\pi} \rho_{K,G}(\theta,t) d\theta = 1$ leads to the following continuity equation: 
\begin{equation}
\frac{\partial\rho_{K,G}}{\partial t}+\frac{\partial}{\partial\theta}\{\rho_{K,G}\left[\omega_0+KR\sin(\Theta-\theta)\right]\}=0.
\label{eq:continuum_equation}
\end{equation}
Next we expand the phase density $\rho_{K,G}(\theta,t)$ in a Fourier series and apply the ansatz by Ott and Antonsen~\cite{ott2008low,pikovsky2015dynamics} to its coefficients to get
\begin{equation}
\rho_{K,G}(\theta,t)=\frac{1}{2\pi}\left(1+\sum_{n=1}^{N}[\alpha_{K,G}(\omega_0,t)]^{n}e^{in\theta}+{\rm c.c}\right), 
\label{eq:dist_exp_ansatz}
\end{equation}
where $\alpha_{K,G}(t) \equiv \alpha(K,G,t)$, and c.c. stands for the complex conjugate. Substituting Eq.~(\ref{eq:dist_exp_ansatz}) into Eq.~(\ref{eq:continuum_equation}) yields 
\begin{equation}
\dot{\alpha}_{K,G}+i\omega_0\alpha_{K,G}+\frac{K}{2}(\alpha_{K,G}^{2}R-R^{*})=0,
\label{eq:alphasKG}
\end{equation}
Inserting Eq.~(\ref{eq:continuum_equation}) into Eq.~(\ref{eq:rKG}), we have that the local order parameters become $z_{K,G} = \alpha_{K,G}^*$. Hence, 
for a general distribution of coupling strengths $P(K,G)$, we get 
\begin{equation}
\begin{aligned}
\dot{r}_{K,G}  &=  \frac{K}{2}(1-r_{K,G}^{2})\langle\langle G'r_{K',G'}\cos(\psi_{K,G}-\psi_{K',G'}\rangle\rangle \\
\dot{\psi}_{K,G}  &=\omega_0 -  \frac{K}{2}(r_{K,G}+r_{K,G}^{-1})\langle\langle G'r_{K',G'}\sin(\psi_{K',G'}-\psi_{K,G})\rangle\rangle 
\end{aligned}
\label{eq:general_reduction}
\end{equation}
where $\langle\langle ... \rangle\rangle=\int \int P(K',G') ... dK'dG'$.

The boundaries of the asynchronous state ($R =0$) can be obtained straightforwardly for arbitrary distributions $P(K,G)$. By considering small 
perturbations $\delta r_{K,G}$ around the incoherent state $r_{K,G}=0$, and setting $\psi_{K,G}=0$, without loss of generality, we get from Eq.~(\ref{eq:general_reduction}) 
\begin{equation}
\dot{\delta r}_{K,G} =  \frac{K}{2}\langle\langle G'\delta r_{K',G'} \rangle\rangle.
\label{eq:deltar}
\end{equation}
By multiplying the previous equation by $G$ and averaging over the distribution $P(K,G)$, we rewrite Eq.~(\ref{eq:deltar}) in terms of a perturbation to the global order
parameter $\delta R = \langle \langle G'\delta r_{K',G'} \rangle \rangle$, $\dot{\delta R} = [ \langle \langle KG \rangle \rangle /2]\delta R$, which 
leads to the critical condition
\begin{equation}
\langle \langle KG \rangle \rangle = 0. 
\label{eq:critical_condition_incoherent}
\end{equation}
Therefore, for $\langle \langle KG \rangle \rangle < 0 $, the oscillators remain incoherent, while for $\langle \langle KG \rangle \rangle > 0$ the incoherent state loses stability, and a partially synchronized state sets in. Notice that Eq.~(\ref{eq:critical_condition_incoherent}) is similar to the condition obtained in Ref.~\cite{sonnenschein2015collective} for identical oscillators subjected to Gaussian noise.

\begin{figure}[t!]
	\centering
	\includegraphics[width=1.0\columnwidth]{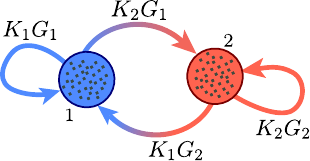}
	\caption{Schematic illustration of a finite system composed of two intertwined subpopulations. Oscillators belonging to subpopulation 1 (2) interact among themselves via couplings $K_1 G_1$ $(K_2 G_2)$, while oscillators from different subpopulations interact asymmetrically with effective couplings $K_1 G_2$ and $K_2 G_1$, as depicted above.}%
	\label{fig1}%
\end{figure} 
 
Let us consider now the case in which the oscillators are coarse-grained into $n$ intertwined subpopulations with joint distribution of couplings given by $P(K,G) = \frac{1}{n}\sum_{q=1}^n  \delta [(K,G) - (K_q,G_q)]$. Substituting the previous expression for $P(K,G)$ into Eqs.~(\ref{eq:general_reduction}) yields
\begin{equation}
\begin{aligned}
\dot{r}_{q}&=-\frac{K_{q}}{2n}(1-r_{q}^{2})\sum_{p=1}^{n}G_{p}r_{p}\cos(\psi_{p}-\psi_{q})\\
\dot{\psi}_{q}&=\omega_0-\frac{K_q}{2n}(r_{q}+r_{q}^{-1})\sum_{p=1}^{n}G_{p}r_{p}\sin(\psi_{p}-\psi_{q}).
\end{aligned}
\label{eq:n_pops}
\end{equation}
In what follows we investigate a special case of the above system, namely, the setup of two intertwined subpopulations~\cite{sonnenschein2015collective}. In this case, we have $n=2$, and Eqs.~(\ref{eq:n_pops}) are reduced to
\begin{equation}
\begin{aligned}
\dot{r}_{1}&=\frac{K_1}{4}(1-r_{1}^{2})[r_1G_{1}+r_{2}G_{2}\cos\delta],\\
\dot{r}_{2}&=\frac{K_2}{4}(1-r_{2}^{2})[r_2G_{2}+r_{1}G_{1}\cos\delta],\\
\dot{\delta}&=-\frac{\sin\delta}{4}\left[(r_{1}+r_{1}^{-1})r_{2}K_{1}G_{2}+(r_{2}+r_{2}^{-1})r_{1}K_{2}G_{1}\right], 
\end{aligned}
\label{eq:2_pop_reduced_system}
\end{equation}
where we have defined the phase-lag $\delta = \psi_1 - \psi_2$. An illustration of a finite network with two intertwined subpopulations can be seen in Fig.~\ref{fig1}. We measure the global synchronization with the classical Kuramoto order parameter as:
\begin{equation}
r(t) e^{i\Phi(t)} = \frac{1}{2}[r_1(t) e^{i\psi_1(t)} + r_2(t) e^{i\psi_2(t)}]
\label{eq:classical_K_op}
\end{equation}
Note that $r(t) e^{i\Phi (t)}$ in the above equation is different from 
the ``weighted'' order parameter $R e^{i\Theta} = \frac{1}{2}[r_1(t) G_1 e^{i\psi_1(t)} + r_2(t) G_2 e^{i\psi_2(t)}]$ [Eq.~(\ref{eq:total_order_parameter})], which can be larger than one.

Equations (\ref{eq:2_pop_reduced_system}) are very similar to the set of equations for identical oscillators under the influence stochastic fluctuations obtained via Gaussian approximation~\cite{sonnenschein2013approximate,sonnenschein2015collective}. 
Actually, the only difference between the reduced system obtained in Ref.~\cite{sonnenschein2015collective} and Eq.~(\ref{eq:2_pop_reduced_system}) is that the former exhibits terms $r_{1,2}^4$ instead of $r_{1,2}^2$ in the equations for $\dot{r}_{1,2}$; and terms proportional to $(r_{1,2}^{-1} + r_{1,2}^3)$ in place of $(r_{1,2}^{-1} + r_{1,2})$ in the equation for $\dot{\delta}$. Notice also that Eqs.~(\ref{eq:2_pop_reduced_system}) could be obtained 
via the Watanabe-Strogatz theory~\cite{watanabe1994constants,pikovsky2008partially} under uniform distribution of constants of motion (Ott-Antonsen manifold)~\cite{hong2011conformists,pikovsky2015dynamics}. 

\begin{figure*}[t!]
	\centering
	\includegraphics[width=2.0\columnwidth]{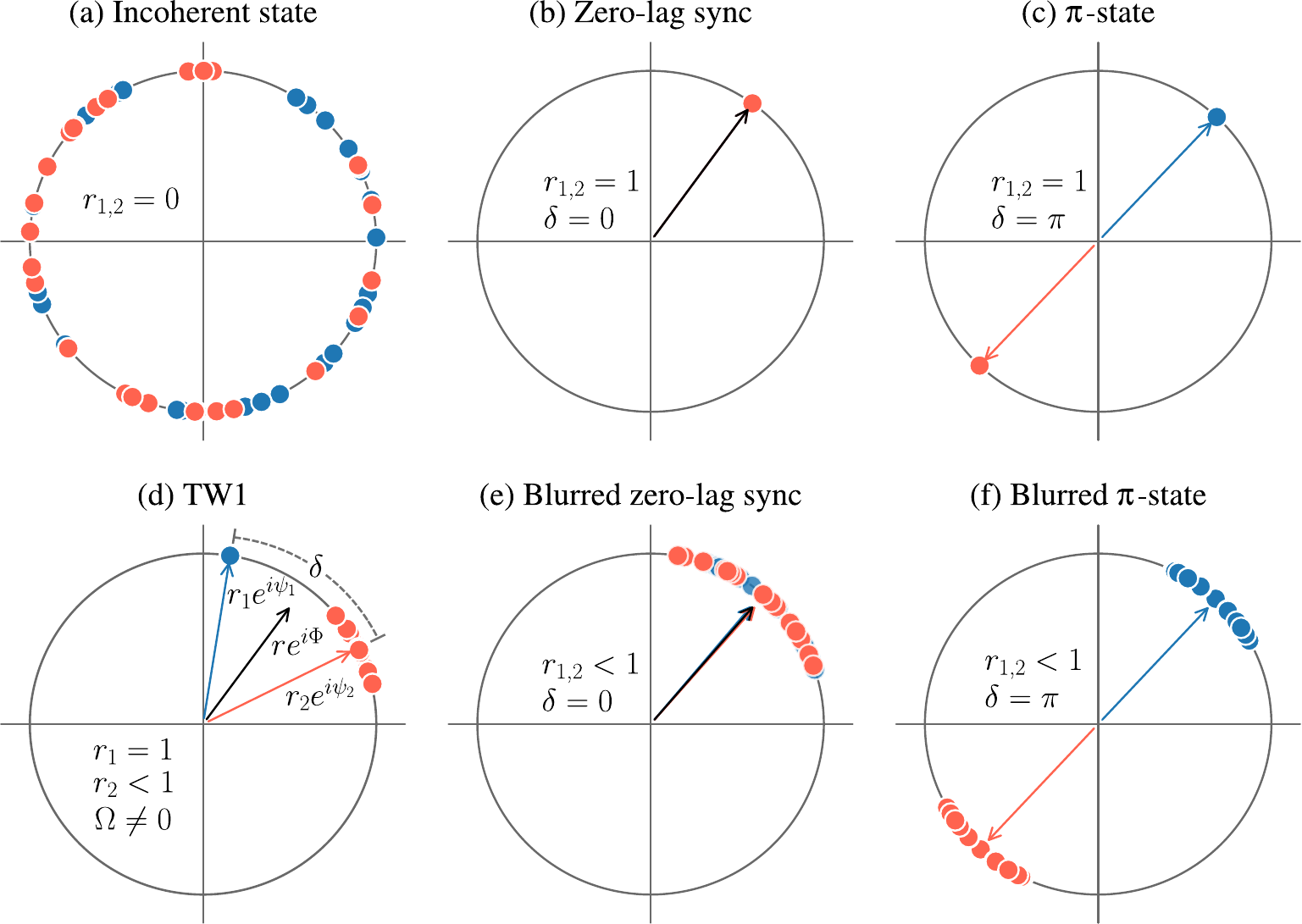}
	\caption{Long-time profile on the complex unit circle of the states observed for a finite system [Eq.~(\ref{eq:KGmodel})] with two intertwined subpopulations coupled as depicted in Fig.~\ref{fig1}: (a) Incoherent state, (b) Zero-lag sync, (c) $\pi$-state, (d) TW1, (e) blurred zero-lag sync, and (f) blurred $\pi$-state. Only in the TW1 state [panel (d)], the collective frequencies $\dot{\Theta}$ [Eq.~(\ref{eq:dotpsi12})], $\dot{\Phi}$ [Eq.~(\ref{eq:classical_K_op})] and $\Omega$ [Eq.~(\ref{eq:avg_freq_OMEGA})] are different from zero, and the oscillators travel across all possible phase values in the co-rotating frame defined by the natural frequency. The configuration of the TW2 state is obtained by interchanging sub-indexes and colors in panel (d).}%
	\label{fig2}%
\end{figure*}

%Given the similarities between these models, we then naturally expect to observe similar dynamical patterns as those reported in~\cite{sonnenschein2015collective}, namely
From Eqs.~(\ref{eq:2_pop_reduced_system}), we expect to observe the following stationary states for the two subpopulation system (see the illustration in Fig.~\ref{fig2}):   (i) the classical incoherent state in which $r=r_{1,2}=0$; (ii) the perfectly synchronized state in which $r_{1,2}=1$ and $\delta=0$ (we denominate this state as ``zero-lag sync''~state); (iii) partially synchronized states characterized by $r_{1,2}<1$ and $\delta = 0$, and which we refer to as ``blurred zero-lag sync'' states; (iv) the so-called ``$\pi$-state'' for which the subpopulations are perfectly synchronized ($r_{1,2}=1$), while remaining diametrically opposed in the phase space ($\delta=\pi$), yielding, hence, a vanishing global synchronization ($r=0$); (v) ``blurred`` $\pi$-states, in which at least one of the subpopulations is partially synchronized ($r_{1,2}<1$) and the peaks of their phase distributions are separated by $\delta = \pi$; and, finally, (vi) the traveling-wave (TW) state~\cite{hong2011kuramoto,petkoski2013mean,sonnenschein2015collective} in which the subpopulations can be either partially or fully synchronized, $0 < r_{1,2}\leq 1$, while keeping a constant phase-lag separation within $0 < \delta < \pi$. The interesting feature of this state is that, in contrast to standard formulations of the Kuramoto model, the oscillators no longer rotate with a common frequency given by the frequency $\omega_0$ of the co-rotating frame--or $\bar{\omega} = \int \omega g(\omega) d\omega$ in the case of non-identical oscillators, where $g(\omega)$ is a frequency distribution~\cite{hong2011kuramoto,petkoski2013mean}--; instead, they settle on a stationary new rhythm whose magnitude will also depend on the coupling parameters. Deviations from the mean frequency $\bar{\omega}$ can be calculated either by the average (or mean-ensemble) frequency
\begin{equation}
\Omega = \frac{1}{N} \sum_{j=1}^N \langle \dot{\theta}_j \rangle_t, 
\label{eq:avg_freq_OMEGA}
\end{equation}
where $\langle ... \rangle_t$ denotes a long-time average; or by the locking frequency $\dot{\Theta}$, defined in Eq.~(\ref{eq:total_order_parameter}) [see also Eq.~(\ref{eq:dotpsi12})]. TW states typically appear in phase-oscillator systems when certain symmetry patterns are broken in the model, such as by the presence of a phase frustration in the sine coupling term~\cite{sakaguchi1986soluble}, asymmetric coupling strength distributions~\cite{hong2011kuramoto,hong2011conformists}, or by natural frequencies asymmetrically distributed~\cite{petkoski2013mean}.

\section{Bifurcation analysis of the reduced system with two subpopulations}
\label{sec:bifurcation_analysis}

For convenience, we adopt the following parametrization for the couplings:
\begin{equation}
K_{1,2} = K_0 \pm \frac{\Delta K}{2} \textrm{ and }G_{1,2} = G_0 \pm \frac{\Delta G}{2}, 
\label{eq:parametrization}
\end{equation} 
where $K_0$ and $G_0$ are the average in- and out-coupling strengths, respectively; parameters $\Delta K$ and $\Delta G$ are defined as the corresponding
coupling mismatches. In our calculations we always consider positive mismatches ($\Delta K, \Delta G>0$). Therefore, if $|K_0| < \Delta K/2$ or $|G_0|< \Delta G/2$, half of the couplings are positive (attractive) and half are negative (repulsive). If one of these conditions is satisfied, we say that the oscillators interact via \textit{mixed couplings}.

By setting $\dot{\delta}=0$, we uncover two possible fixed-point solutions for phase-lag $\delta$:
\begin{eqnarray}
\delta & = & m\pi\mbox{, }m\in\mathbb{Z}\label{eq:pi_state_solution}\\
0 & = & (r_{1}+r_{1}^{-1})r_{2}K_{1}G_{2}+(r_{2}+r_{2}^{-1})r_{1}K_{2}G_{1}\label{eq:tw_state_solution}.
\end{eqnarray} 
Equation~(\ref{eq:pi_state_solution}) corresponds to the solution of partially synchronized states with no separation between populations (even $m$) and $\pi$-states (odd $m$), while Eq.~(\ref{eq:tw_state_solution}) gives the condition for the existence of TW states. We can verify that TWs appear only for $0 < \delta < \pi $ by rewriting the equations for $\dot{\psi}_{1,2}$ together with Eq.~(\ref{eq:tw_state_solution}) as: 
\begin{equation}
\lim_{t\rightarrow\infty}\dot{\psi}_{1,2}=\lim_{t\rightarrow\infty}\dot{\Theta}=\omega_0 - \sin\delta\frac{r_{2}+r_{2}^{-1}}{4}K_{2}G_{1}r_{1}.
\label{eq:dotpsi12}
\end{equation}
Therefore, spontaneous drifts in the collective frequencies occur only for intermediate values of the phase-lag $\delta$; otherwise, for $\delta=m\pi$, oscillators rotate with collective frequencies $\dot{\psi}_{1,2}=\omega_0$, which here is set  to $\omega_0=0$.

From the parametrization in Eq.~(\ref{eq:parametrization}), we realize the critical conditions $K_{1,2}=0$, or
\begin{equation}
K_{0} = \pm \frac{\Delta K}{2}.
\label{eq:K0c1_DK_2}
\end{equation}
When one of the above conditions holds, it follows that 
one subpopulation is deprived of receiving inputs from other
nodes (including from the same subpopulation), and its oscillators have instead only out-couplings towards nodes external to their subpopulation. Similarly, if one of the out-couplings vanishes, $G_{1,2}=0$, the corresponding subpopulation ceases to influence the dynamics of the rest of the network and starts acting only as link receiver. As we shall see, these conditions play an important role in the phase diagram of Eq.~(\ref{eq:2_pop_reduced_system}). In the sequel, we calculate the coupling ranges in which the states discussed
in the previous section appear.

\subsection{Incoherent state}

The first critical condition of Eqs.~(\ref{eq:2_pop_reduced_system}) is given by the stability 
analysis of the incoherent state performed in the last section. For the case of two subpopulations with $P(K,G)=\frac{1}{2}\delta[(K,G)-(K_1,G_1)]+\frac{1}{2}\delta[(K,G)-(K_2,G_2)]$, Eq.~(\ref{eq:critical_condition_incoherent}) reads $K_1 G_1 + K_2 G_2 = 0$, and by solving it in terms of the average in-coupling strength we have
\begin{equation}
K_0 = - \frac{\Delta K \Delta G}{4G_0}.
\label{eq:K0c_incoherent}
\end{equation}
The above equation, therefore, delineates the boundary of the incoherent state.

\begin{figure*}[t!]
	\centering
	\includegraphics[width=2.0\columnwidth]{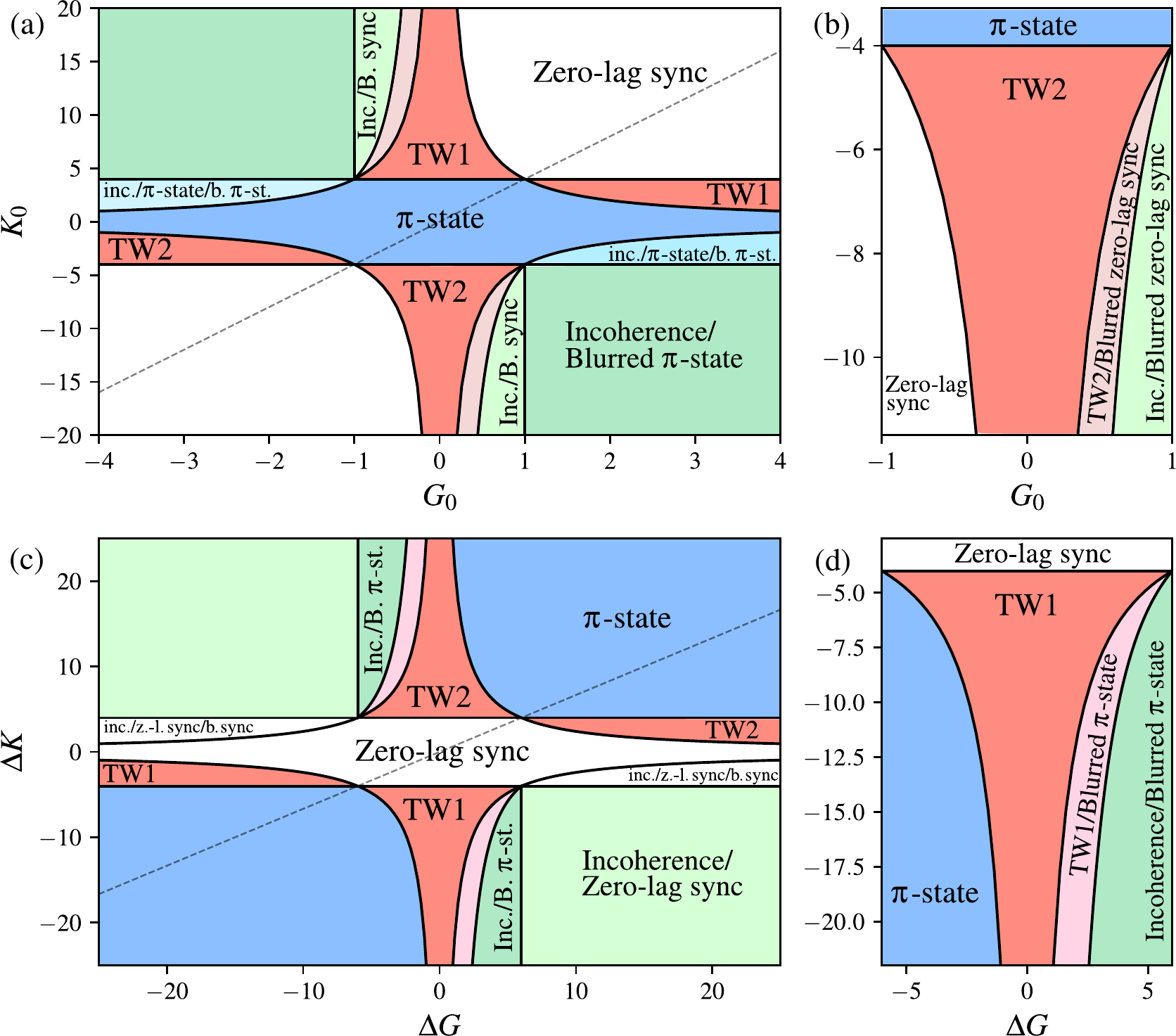}
	\caption{Bifurcation diagram of the reduced system [Eq.~(\ref{eq:2_pop_reduced_system})] for (a,b) $\Delta K = 8$ and $\Delta G = 2$; and (c,d) $K_0 = 3$ and $G_0 = 2$ . Zero-lag sync corresponds to the perfectly synchronized state ($r_{1,2} = 1$ and $\delta = 0$). ``$\pi$-state'' refers to the state in which $r_{1,2}=1$ and phase-lag separation $\delta = \pi$. Similarly, blurred zero-lag and blurred $\pi$-states denote the states with $r_{1,2}<1$ along with $\delta =0$ and $\delta = \pi$, respectively.
Traveling wave states are characterized by $r_1 = 1$, $r_2 < 1$ (TW1), and $r_2 = 1$, $r_1 < 1$ (TW2). Both TW1 and TW2 exhibit $\Omega \neq 0$ [Eq.~(\ref{eq:avg_freq_OMEGA})].  Solid lines are obtained from Eqs.~(\ref{eq:K0c_incoherent})-(\ref{eq:partially_sync_stab_cond_blurred}) and Eqs.~(\ref{eq:TW1_conditions_K0})-(\ref{eq:TW2_conditions_K0}). Dashed line in panels (a) and (c) depict the condition $K_0 \Delta G - \Delta K G_0 = 0$ for which the couplings are symmetric, i.e., when the network connections are undirected. Panels (b) and (d) show zoomed-in regions of panels (a) and (c), respectively.}%
	\label{fig3}%
\end{figure*}

\subsection{$\pi$-states}

For the $\pi$-state, the fixed point solutions read $r_{1,2}=1$ and $\delta = \pi$. Linear stability reveals that the $\pi$-state is stable
for average in-couplings given by 
\begin{equation}
\begin{aligned}
-\frac{\Delta K}{2}< K_0 < \min\left\{ \frac{\Delta K}{2}, \frac{\Delta K \Delta G}{4 G_0} \right\}, \mbox{ for }G_0>0;\\
\max\left\{-\frac{\Delta K}{2}, \frac{\Delta K \Delta G}{4 G_0} \right\}< K_0 < \frac{\Delta K}{2}, \mbox{ for }G_0<0.
\end{aligned}
\end{equation}

The other state characterized by subpopulations diametrically opposed is the blurred $\pi$-state, and its fixed points are defined by 
$r_1 G_1 = r_2 G_2$ and $\delta = \pi$. Linearizing the dynamics 
about this state, we find that the corresponding Jacobian matrix has a single nonzero eigenvalue, $\lambda = K_1 G_1 (1-r_1^2)/4 + K_2 G_2 (1-r_2^2)/4$. Hence, and because couplings $G_1$ and $G_2$ must have the same sign so that $r_1 G_1 = r_2 G_2$ is a physical solution, we have that blurred $\pi$-states appear when
\begin{equation}
K_0 < \frac{\Delta K \Delta G}{4 G_0},\mbox{ for }|G_0|>\frac{\Delta G}{2}.
\label{eq:blurred_pi_state}
\end{equation}
Thus, both $\pi$-states with fully synchronous subpopulations ($r_{1,2}=1$) and with partial synchronization ($r_{1,2}<1$) are yielded by systems described by Eqs.~(\ref{eq:2_pop_reduced_system}). Observe also that $r_1 G_1 = r_2 G_2$ defines a one-parameter family of fixed points. 
Furthermore, from Eq.~(\ref{eq:blurred_pi_state}) we notice that incoherent and blurred $\pi$-states may coexist in a large region of the parameter space defined by coupling strengths $G_0$ and $K_0$.

\subsection{Zero-lag sync and partially synchronized states}

For the zero-lag sync state, we linearize Eqs.~(\ref{eq:2_pop_reduced_system}) around $r_{1,2}=1$ and $\delta = 0$, and seek a zero eigenvalue of the related Jacobian matrix. By following this calculation, we find that the zero-lag sync state emerges for average in-coupling strengths given by 
\begin{equation}
\begin{aligned}
K_0 &> \max\left\{ \frac{\Delta K}{2},\frac{\Delta K \Delta G}{4 G_0}\right\}, 
\mbox{ for }G_0>0;\\
K_0 &< \min\left\{-\frac{\Delta K}{2}, \frac{\Delta K \Delta G}{4 G_0} 
\right\}, \mbox{ for }G_0<0.
\end{aligned}
\label{eq:zero_lag_sync_stab_cond}
\end{equation} 
Next, by linearizing Eqs.~(\ref{eq:2_pop_reduced_system}) around an arbitrary partially synchronized solution ($r_1 G_1 = -r_2 G_2$ and $\delta=0$), we find that such states, which we have denominated blurred zero-lag sync states, must occur for parameters in the range
\begin{equation}
K_0 <  \frac{\Delta K \Delta G}{4 G_0}, \textrm{ for }|G_0| < \frac{\Delta G}{2}.
\label{eq:partially_sync_stab_cond_blurred}
\end{equation}
Any state that satisfies $r_1 G_1 = -r_2 G_2$ and $\delta=0$ is a fixed point of Eqs.~(\ref{eq:2_pop_reduced_system}). From this we foresee that several partially synchronized states should coexist
in the region delimited by Eq.~(\ref{eq:partially_sync_stab_cond_blurred}).
As in the case of blurred $\pi$-states, blurred zero-lag sync solutions yield a single nonzero Jacobian eigenvalue, $\lambda = K_1 G_1 (1-r_1^2)/4 + K_2 G_2 (1-r_2^2)/4$; hence, the latter states do not coexist with zero-lag sync nor with $\pi$-states, because in regions where $r_{1,2}=1$ we have $\lambda = 0$, and the blurred zero-lag sync states lose stability. Notice further that in order to $r_1 G_1 = - r_2 G_2$ be a physical solution, out-couplings $G_1$ and $G_2$ must have opposite signs; thus, partially synchronized
states with $\delta =0$ are expected to appear only in the presence of mixed-out coupling strengths, i.e., for $|G_0| < \Delta G/2$, as indicated in Eq.~(\ref{eq:partially_sync_stab_cond_blurred}).

\subsection{Traveling waves}

We now turn our attention to the stationary TW states. Numerical results show us that two possible TW states are
manifested by the system (\ref{eq:2_pop_reduced_system}). In the first state, which we refer to as ``TW1'', the first subpopulation remains fully synchronized ($r_1=1$), whereas the second one exhibits partial synchronization ($r_2<1$). We label as ``TW2'' the opposite situation, i.e., when $r_2=1$ and $r_1 < 1$. In both 
states we have $0 < \delta < \pi$ and $\Omega > 0$. By setting $r_1=1$ in Eqs.~(\ref{eq:2_pop_reduced_system}), we find the following fixed point solutions for 
$r_2$ and $\delta$:
\begin{equation}
r_{2}=\sqrt{-\frac{K_{2}G_{1}}{K_{2}G_{1}+2K_{1}G_{2}}}\textrm{ and }\cos\delta = - \frac{G_2}{G_1}r_2.
\label{eq:order_parameter_1_TW}
\end{equation} 
Since $0 < r_2 < 1$, we have that the solution of the TW1 exists for 
\begin{equation}
-K_1 G_2 < K_2 G_1 < 0. 
\label{eq:TW1_condition}
\end{equation}
By writing Eq.~(\ref{eq:TW1_condition}) in terms of the parametrization in Eq.~(\ref{eq:parametrization}) and considering $\Delta K, \Delta G > 0$, we find that the regions with TW1 are outlined by the following conditions: 
\begin{equation}
\begin{aligned}
\frac{\Delta K \Delta G}{4 G_0} <& K_0 < \frac{\Delta K}{2}, \textrm{ for }G_0> \frac{\Delta G}{2}.
\end{aligned}
\label{eq:TW1_conditions_K0}
\end{equation}
The linearization of Eqs.~(\ref{eq:2_pop_reduced_system}) about the fixed points of Eq.~(\ref{eq:order_parameter_1_TW}) also reveals a second region in which the TW1 state is stable:  
\begin{equation}
\begin{aligned}
 \frac{\Delta K}{2} < K_0 < -\frac{\Delta K}{8} \left[ \frac{\Delta G}{G_0}\right]^2,  &\textrm{ for } -\frac{\Delta G}{2}<G_0< 0;\\
  \frac{\Delta K}{2} < K_0 < \frac{\Delta K \Delta G}{4 G_0},  &\textrm{ for } 0 <G_0< \frac{\Delta G}{2}.
 \end{aligned}
\label{eq:TW1_condition_K0_2}
\end{equation}
The solutions for $r_1$ and $\delta$, and the critical conditions for the TW2 state are obtained by interchanging indexes ``1'' and ``2'' in Eqs.~(\ref{eq:order_parameter_1_TW}) and~(\ref{eq:TW1_condition}). By following the same procedure for 
the corresponding TW2 solutions, one uncovers that this state 
appears for average in-coupling strengths given by
\begin{equation}
\begin{aligned}
- \frac{\Delta K}{2} < K_0 <  \frac{\Delta K \Delta G}{4 G_0}, &\textrm{ for }G_0< - \frac{\Delta G}{2};\\
\frac{\Delta K \Delta G}{4 G_0} < K_0 < -\frac{\Delta K}{2}, &\textrm{ for } -\frac{\Delta G}{2} < G_0 < 0;\\
\frac{\Delta K}{8} \left[ \frac{\Delta G}{G_0}\right]^2   < K_0 < -\frac{\Delta K}{2},  &\textrm{ for } 0 <G_0< \frac{\Delta G}{2}.
\end{aligned}
\label{eq:TW2_conditions_K0}
\end{equation}
Alternatively, the conditions of the TW2 state [Eq.~(\ref{eq:TW2_conditions_K0})] could be obtained by letting $(\Delta K,\Delta G) \rightarrow (-\Delta K, -\Delta G)$ in the TW1 conditions [Eq.~(\ref{eq:TW1_conditions_K0}) and~(\ref{eq:TW1_condition_K0_2})]. 

\begin{figure*}[t!]
	\centering
	\includegraphics[width=2.0\columnwidth]{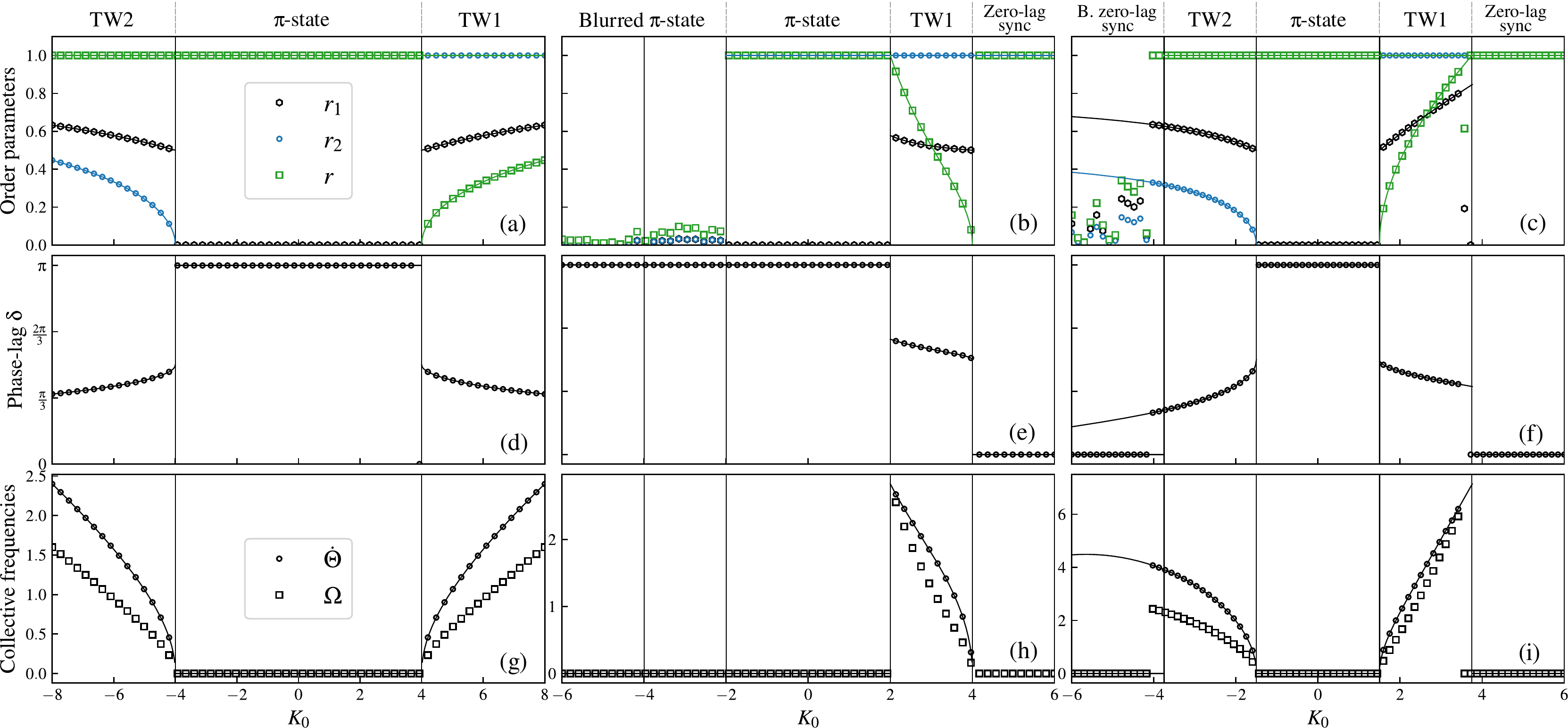}
	\caption{Order parameters $r_{1,2}$, $r$, phase-lag $\delta$, and collective frequencies $\dot{\Theta}$ [Eq.~(\ref{eq:dotpsi12})] and $\Omega$ [Eq.~(\ref{eq:avg_freq_OMEGA})] for (a,d,h) $G_0 = 0$, $\Delta K = 8$, and $\Delta G = 2$; (b,e,h) $G_0 = 2$, $\Delta K = 8$, and $\Delta G = 2$; and (c,f,i)  $G_0 = 2$, $\Delta K = 3$, and $\Delta K = 10$ and $\Delta G = 2$. Dots are obtained by numerically integrating the original system [Eq.~(\ref{eq:KGmodel})] using the Heun's method with $N=10^4$ oscillators. For each coupling value $K_0$, the quantities are averaged over $t \in [500,1500]$ with a time-step $dt=0.005$. In all panels, initial conditions $\theta_i (t=0)$ $\forall i$ are randomly distributed according to a uniform distribution between $[-\pi,\pi]$. Solid lines correspond to the analytic solutions obtained in Sec.~\ref{sec:bifurcation_analysis}.}%
	\label{fig4}%
\end{figure*}

Figure~\ref{fig3}(a) depicts the bifurcation diagram outlined 
by the critical conditions in Eqs.~(\ref{eq:K0c_incoherent})-(\ref{eq:partially_sync_stab_cond_blurred}) and Eqs.~(\ref{eq:TW1_conditions_K0})-(\ref{eq:TW2_conditions_K0}).
As can be seen, for high values of both $K_0$ and $G_0$, the dynamics converges either to perfect synchronization or to incoherence/blurred $\pi$-state; for intermediate values, however, multiple 
regions of bistability appear. Interestingly, although simpler, the model in Eq.~(\ref{eq:KGmodel}) exhibits a more complex dynamics than its stochastic version~\cite{sonnenschein2015collective}. By comparing the diagrams of Fig.~\ref{fig3} with their counterparts in Ref.~\cite{sonnenschein2015collective}, we see that the regions with coexistence between different states become proportionally larger when noise is absent. A similar effect was observed by Hong and Strogatz when comparing the findings in Refs.~\cite{hong2011kuramoto,hong2011conformists}. 

By rewriting the critical couplings in Eqs.~(\ref{eq:K0c_incoherent})-(\ref{eq:partially_sync_stab_cond_blurred}) and Eqs.~(\ref{eq:TW1_conditions_K0})-(\ref{eq:TW2_conditions_K0}) in terms of coupling mismatches, we derive the stability 
diagram spanned by the parameters $\Delta G$ and $\Delta K$. Similarly to what was verified in Ref.~\cite{sonnenschein2015collective}, the arrangement of the transitions in Fig.~\ref{fig3} evidences some rules for the occurrence of the collective states manifested by the system (\ref{eq:KGmodel}). First, in order to observe $\pi$-states, mixed in-couplings are required [see that the blue areas in Fig.~\ref{fig3}(c) occur for $\Delta K > 2K_0$]. States TW1 and TW2 emerge when either mixed in- or out-coupling exist, but never when both types of couplings are mixed -- in the latter case, only incoherence and $\pi$-states are possible. Bistability of 
TWs and incoherence with $\pi$-states appear when mixed in-coupling strengths 
exist, whereas we observe bistable regions TW/zero-lag sync and TW/incoherence 
when only out-couplings are mixed. Finally, we emphasize the importance 
of the directness in the network connections for the emergence of traveling waves. For $K_0 \Delta G - \Delta K G_0 = 0$, the coupling strengths connecting the subpopulations become equal and the interaction is no longer asymmetric. By projecting this expression onto the stability 
diagram (see the dashed lines in Fig.~\ref{fig3}), we see that the symmetry condition does not intersect TW regions; therefore, asymmetric interactions are necessary for the 
emergence of such states. Nevertheless, as seen in Fig.~\ref{fig3}, $\pi$-states are crossed by the line imposed by the symmetry relation, meaning that asymmetric couplings are not required for the existence of $\pi$-states.

The diagrams in Fig.~\ref{fig3} also allow us to reexamine the results in Refs.~\cite{hong2011kuramoto,hong2011conformists,hong2012mean} as particular cases of the present
model. As mentioned previously, in Refs.~\cite{hong2011kuramoto,hong2011conformists}, the authors
studied Kuramoto oscillators subjected to attractive and repulsive in-couplings strengths ($\Delta G = 0$ in our notation). In that setting, the couplings are regarded as a property of the nodes; thus, a fraction of the oscillators tends to align with the mean-field (conformists oscillators), while the rest is repelled by it (contrarians oscillators). As shown in Refs.~\cite{hong2011kuramoto,hong2011conformists}, the absence of out-coupling strengths does not impede the
system from reaching $\pi$-states and traveling-waves. Indeed, if we set $\Delta G = 0$ in Fig.~\ref{fig3}(b) and
follow the transitions along the $K_0$-axis, we see that the system switches from TW1 to TW2 via crossing the central $\pi$-state area. In their follow-up study~\cite{hong2012mean}, the couplings were treated
as properties of the links, that is, the coupling terms were placed inside the summation over the neighbors' connections instead of outside as in Refs.~\cite{hong2011kuramoto,hong2011conformists}. This coupling setting is equivalent to the model in Eq.~(\ref{eq:KGmodel}) in the absence of in-coupling strength mismatches ($\Delta K = 0$). Interestingly, despite the presence of mixed couplings, neither traveling waves nor $\pi$-states were detected, but rather only partially synchronized and incoherent states~\cite{hong2012mean}. The fact that mixed out-couplings under no mismatch in the in-couplings yield only a classical mean-field behavior is evident in Fig.~\ref{fig3}, where we see that for $\Delta K = 0$ the only possible state is zero-lag sync.

\section{Simulations}
\label{sec:simulations}

\begin{figure*}[t!]
	\centering
	\includegraphics[width=2.0\columnwidth]{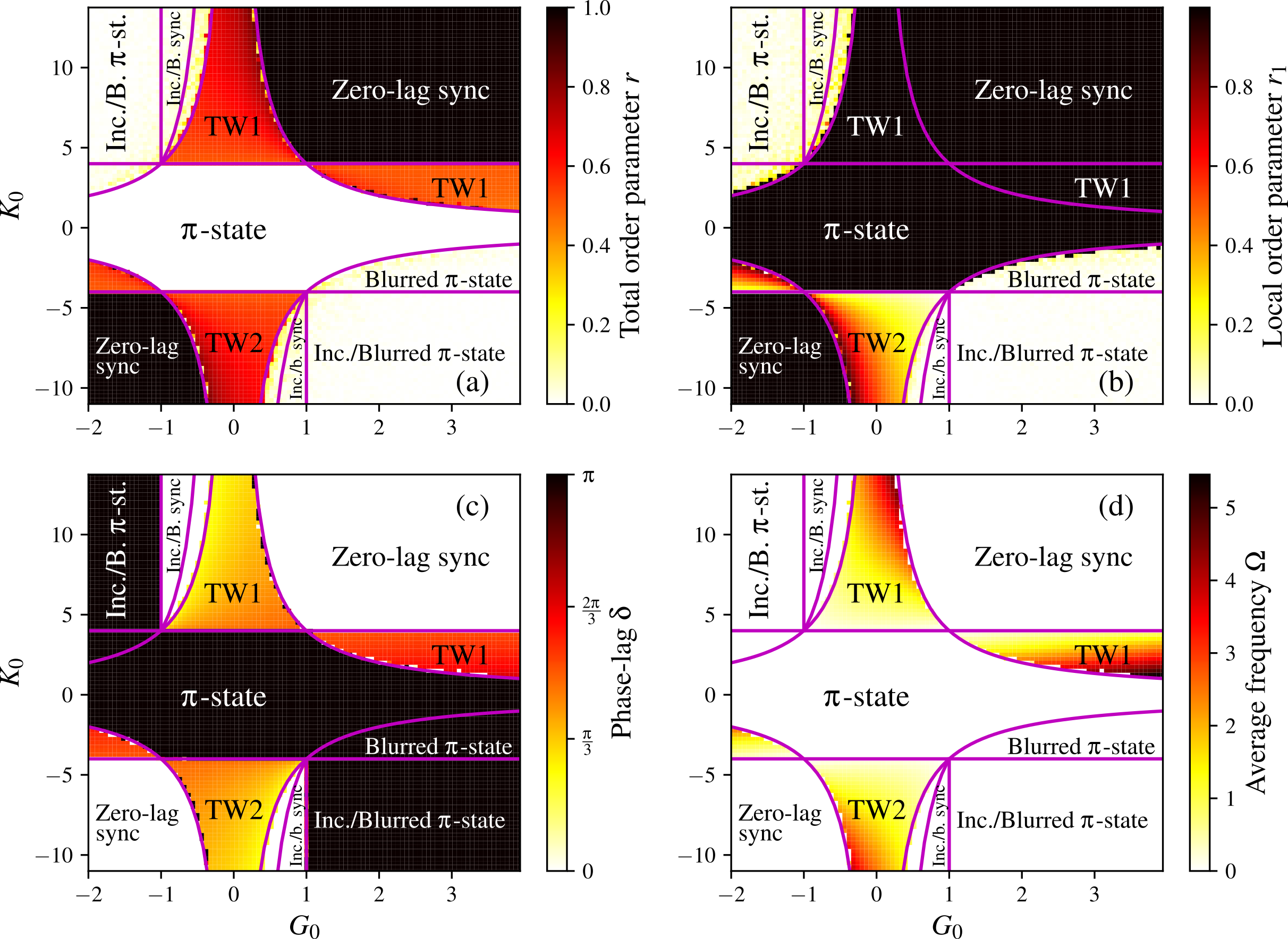}
	\caption{Comparison between simulations (colormaps) and theory (solid lines). (a) Total order parameter $r$; (b) local order parameter $r_1$ of subpopulation 1; (c) phase-lag $\delta$ measuring the separation between the two subpopulations; and (d) average frequency $\Omega$ [Eq.~(\ref{eq:avg_freq_OMEGA})]. For each pair of coupling $(G_0,K_0)$, Eqs.~(\ref{eq:KGmodel}) are evolved numerically with the Heun's method considering $N=10^4$ oscillators and with integration time step $dt=0.005$. The long-time behavior of each parameter is quantified by averaging the trajectories over $t \in [500,1000]$. For all $(G_0,K_0)$, the initial phases $\theta_i (t=0)$ are drawn uniformly at random over the interval $[-\pi,\pi]$. Coupling mismatch parameters: $\Delta K = 8$ and $\Delta G = 2$.}
	\label{fig5}
\end{figure*}

Let us now compare the results of the bifurcation analysis in the previous section with numerical simulations of the finite original dynamics [Eq.~(\ref{eq:KGmodel})]. Figure~\ref{fig4} shows the evolution of order parameters, phase-lag separation $\delta$, and collective frequencies, $\Omega$ and $\dot{\Theta}$, as a function of $K_0$ for different choices of $G_0$. All the simulations are performed by integrating Eqs.~(\ref{eq:KGmodel}) with the Heun's method with a time step $dt=0.005$ and considering total number of oscillators $N=10^4$ (see the caption of Fig.~\ref{fig4} for more details). For $G_0=0$ [Fig.~\ref{fig4} (a), (d), and (g)] we observe that
the system transitions from TW2 to $\pi$-state, and then subsequently to TW1, 
as correctly predicted by the critical conditions depicted in the diagram of Fig.~\ref{fig3}. In Fig.~\ref{fig4}(b), we see that at $K_0 = -2$ the subpopulations abruptly synchronize as they switch from blurred $\pi$-states to $\pi$-state. A similar discontinuous transition of the local order parameters $r_{1,2}$
was observed for similar parameter configurations in the stochastic version of the system (\ref{eq:KGmodel})~\cite{sonnenschein2015collective}. Abrupt transitions 
are also seen in Fig.~\ref{fig4}(c), but this time as a consequence of the transition from blurred zero-lag sync to TW2 state. In Fig.~\ref{fig4}(c) we also observe irregular points in the  ``B. zero-lag sync'' region. Those points correspond to partially synchronized states and display such an irregular pattern because of the one-parameter family of solutions that exists in that region; specifically, different initial conditions drive the system to different stationary states that satisfy  $r_1 G_1 = - r_2 G_2$.
The solid branches in ``B. zero-lag sync'' area correspond to TW2 solutions, which are also stable for $K_0 \lesssim -3.9$ in Fig.~\ref{fig4}(c) [see also Fig.~\ref{fig3}(b)], but are not  obtained numerically with the initial conditions used in Fig.~\ref{fig4}. We shall return to this point shortly.

Notice in Fig.~\ref{fig4}(g)-(i) that $|\Omega| \leq |\dot{\Theta}|$. The reason for this resides in the fact that
$\Omega$ is a microscopic average of the instantaneous frequencies $\langle \dot{\theta}_i \rangle$, while $\dot{\Theta}$ [and equivalently $\dot{\Phi}$ in Eq.~(\ref{eq:classical_K_op})] quantifies how
fast the center of the bulk formed by entrained oscillators rotates. Therefore, oscillators that are
not locked with the mean-field contribute to the sum in Eq.~(\ref{eq:avg_freq_OMEGA}) with $\langle \dot{\theta}_i \rangle_t \approx 0$, thus reducing the value of $\Omega$ in comparison with its upper bound $\dot{\Theta}$. The latter frequency offers in the present case the slight advantage of being calculated directly from the solutions in Eq.~(\ref{eq:dotpsi12}). Analogously, $\Omega$ can be
estimated analytically (not shown here) through the ensemble average $\Omega = \int_{-\pi}^{\pi} \int \int \dot{\theta}\rho(\theta,t|K,G)dKdGd\theta$. 

\begin{figure}[t!]
	\centering
	\includegraphics[width=1.0\columnwidth]{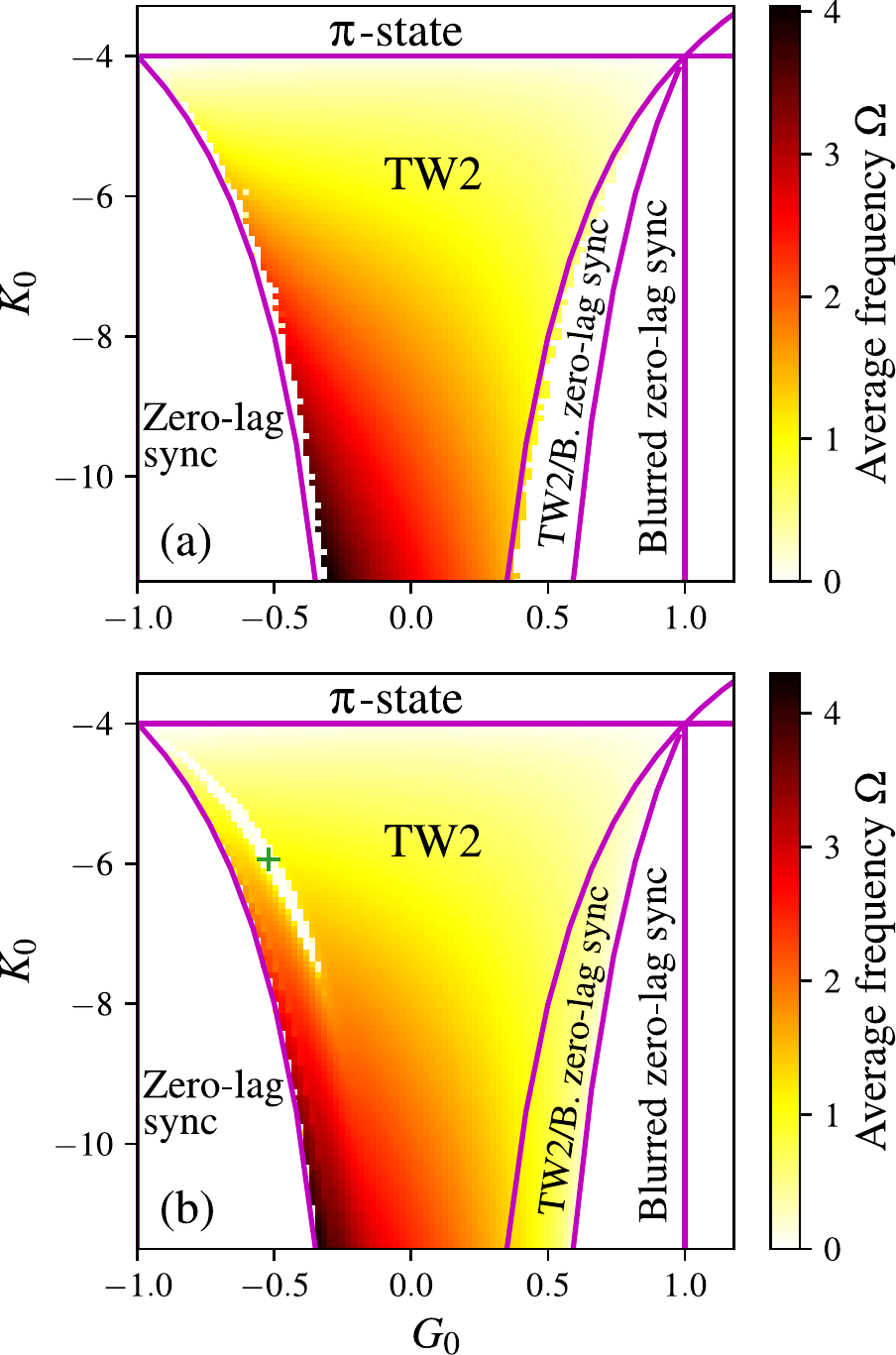}
	\caption{Comparison between simulations (colormaps) and theory (solid lines) for different sets of initial conditions: (a) $\theta_i(t=0)$ randomly chosen from the uniform distribution $[-\pi,\pi]$ (same as in Fig.~\ref{fig5}); (b) for each coupling pair ($G_0,K_0$) the initial phases of subpopulation 1 and 2 were drawn from Gaussian distributions with standard deviation $\sigma=2$, and means $\theta_1$ and $\theta_2$, respectively, which were chosen uniformly at random between $[-\pi,\pi]$. The ``+'' marks a point in the diagram for which the behavior observed in the simulation departs from the dynamics predicted by the theory. Figure~\ref{fig7} shows the temporal evolution of the collective variables at the ``+'' point depicted in panel (b). Other parameters: $N=10^4$, $\Delta K = 8$ and $\Delta G = 2$. In both panels the resolution of the grid is $100\times 100$ couplings. Integration was performed with the Heun's method using time step $dt=0.005$.}%
	\label{fig6}%
\end{figure}

To conclude this section, in Fig.~\ref{fig5} we compare the theoretical results with simulations considering coupling parameters over a $G_0 \times K_0$ grid. Our goal with this approach is to inspect for a larger set of parameters whether the stability analysis performed in the previous section correctly predicts the stability regions shown in Fig.~\ref{fig3}. For each coupling pair $(G_0,K_0)$, Eqs.~(\ref{eq:KGmodel}) are integrated numerically, and the global variables are averaged over $t \in [500,1500]$ with a time step $dt= 0.005$. As can be seen in Fig.~\ref{fig5}, the boundaries of the collective states are predicted very accurately by the theory. Seeking to verify the bistable behavior of the model,  in Fig.~\ref{fig6} we show simulations results for a zoomed region of the space in Fig.~\ref{fig5} considering different initial conditions:  in Fig.~\ref{fig6}(a), phases $\theta_i$ are initiated with values distributed uniformly at random between $[-\pi,\pi]$.  Initial conditions were chosen differently for Fig.~\ref{fig6}(b); specifically, for each point in the grid $G_0 \times K_0$, the phases of populations 1 and 2 were chosen from Gaussian distributions with standard deviation $\sigma = 2$, and means $\theta_1$ and $\theta_2$, which were taken uniformly at random between $[-\pi,\pi]$. By initiating the oscillators in this way, we observe in Fig.~\ref{fig6} that the system converges to TW2 in the region where this state was predicted to coexist with partial synchronization.

\begin{figure}[t!]
	\centering
	\includegraphics[width=1.0\columnwidth]{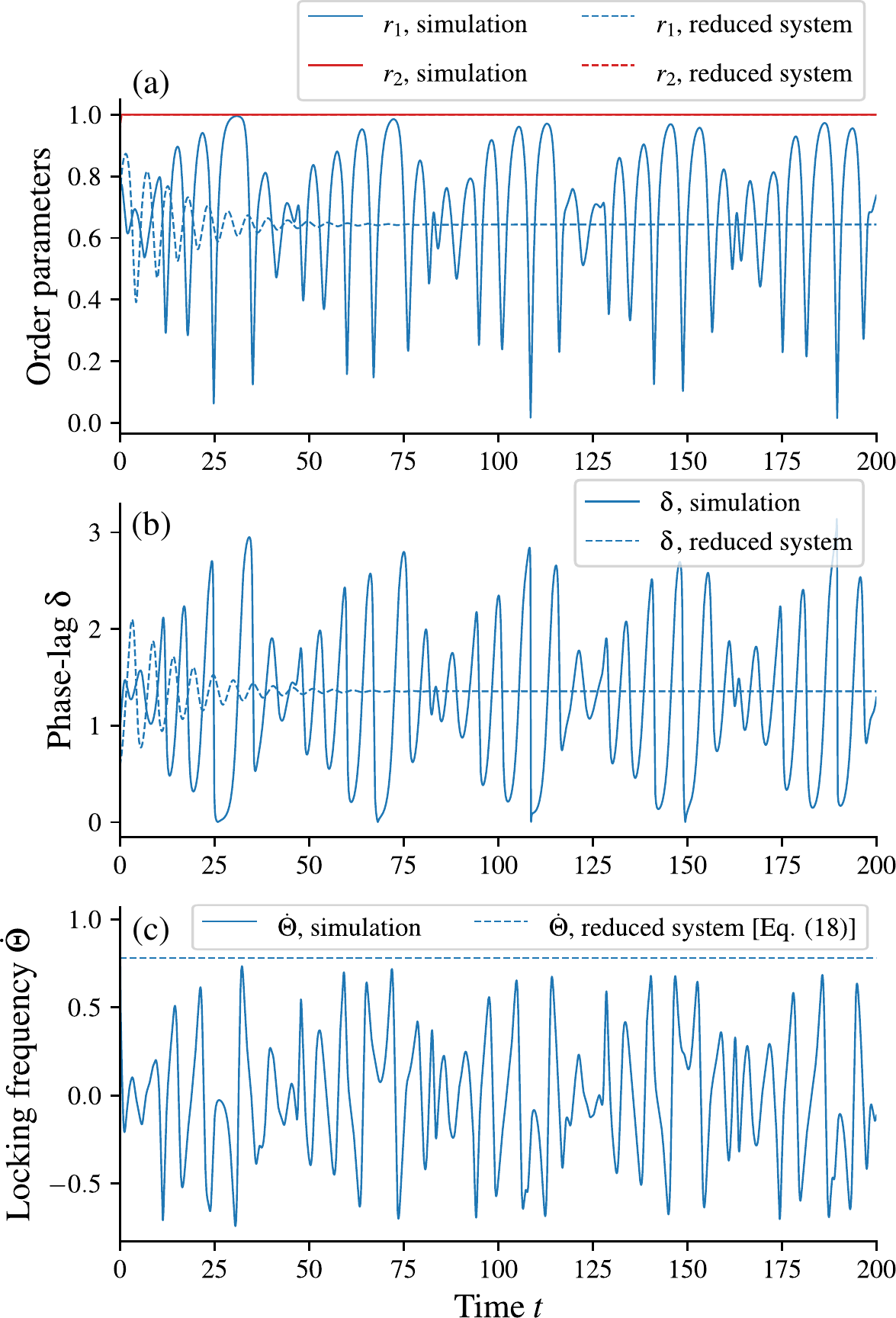}
	\caption{Temporal evolution of (a) local order parameters $r_{1,2}$, (b) phase-lag $\delta$, and (c) locking frequency $\dot{\Theta}$ [Eq.~(\ref{eq:dotpsi12})]. Solid lines are obtained from simulations, while dashed lines correspond to the results yielded by the numerical integration of the reduced system [Eq.~(\ref{eq:2_pop_reduced_system})]. In panel (a) the solid and dashed lines of $r_2$ overlap each other at $r_2 = 1$. Average in- and out-coupling strengths are taken from the ``+'' point in Fig.~\ref{fig6}(b), that is, $(G_0,K_0)=(-0.5,-5.95)$. Other parameters: $N=10^4$ oscillators, $\Delta K = 8$, $\Delta G = 2$, and $dt = 0.005$. }%
	\label{fig7}%
\end{figure}

\section{Accuracy of the Ott-Antonsen reduction}
\label{sec:accuracy}

Although we have observed a good agreement between simulations and the theory for the states previously discussed, there are also dynamical patterns which seem not to be captured by the OA reduction. Figure~\ref{fig6} shows an example: there we observe a set of points with $\dot{\Theta}=0$
in the TW2 area, i.e., a region where one would expect stationary states with $\dot{\Theta} \neq 0$.  By inspecting the temporal trajectories of the local order parameters $r_{1,2}$ in Fig.~\ref{fig7}, we see that such states do have a different nature from traveling waves. The trajectories in Figs.~\ref{fig7} actually resemble \textit{breathing chimera} states~\cite{abrams2008solvable} in which one subpopulation remains fully locked, while the other exhibits an oscillating synchrony. In the figure, we compare 
the time evolution of the original model (\ref{eq:KGmodel}) with the numerical integration 
of the reduced system [Eq.~(\ref{eq:2_pop_reduced_system})] using the same initial conditions. As it is seen, while the finite subpopulation 1 shows oscillating synchrony, the solution provided by the theory converges to a constant value. Although chimera states have been studied extensively with the OA reduction, Figs.~\ref{fig6} and~\ref{fig7} suggest that such solutions in the present model might lie outside the OA manifold. Interestingly, Refs.~\cite{hong2011kuramoto,hong2011conformists,sonnenschein2015collective} did not report solutions akin to the ones shown in Fig.~\ref{fig7}.

\section{Conclusion}
\label{sec:conclusions}

In this paper we have studied a variant of the Kuramoto model in which identical oscillators are coupled via in- and out-coupling strengths, which in turn can have positive and negative values. Similarly to the setting considered in Ref.~\cite{sonnenschein2015collective}, heterogeneity in the interactions was introduced by dividing the oscillators into two mutually coupled subpopulations (each one characterized by a distinct pair of couplings), so that connections within the same subpopulation remain symmetric, while connections between subpopulations are asymmetric. In the infinite size limit, we applied the theory by Ott and Antonsen~\cite{ott2008low} to obtain a reduced set of equations. With the reduced description of the original system, we performed a thorough bifurcation analysis whereby a rich dynamical behavior was revealed. We showed that the present system exhibits different types of $\pi$-states and traveling-waves, along with classical incoherent and partially synchronized states. Though the transitions among these states bear some similarity to those uncovered for the model with Gaussian white noise in Ref.~\cite{sonnenschein2015collective}, we have found that our model exhibits a more intricate long-term dynamics than that of observed for its noisy version. The reason for this conclusion resides in the different types of $\pi$- and zero-lag sync states uncovered (which may consist of either perfectly or partially synchronized subpopulations), and in the observation of wider regions in the parameter space displaying bistability. These findings for the seemingly simpler system are in line with previous studies~\cite{hong2011kuramoto,hong2011conformists} comparing the dynamics of identical oscillators with that of non-identical ones. [Although the system in Eq.~(\ref{eq:KGmodel}) and the one of  Ref.~\cite{sonnenschein2015collective} are both models of identical oscillators, the inclusion of Gaussian white noise yields equivalent phenomenology--with respect to the linearized dynamics--to the case of phase oscillators with natural frequencies drawn from Lorentzian distributions; see the discussion in Ref.~\cite{pietras2016equivalence}.]    

Despite the excellent agreement between simulations and theory, for a small set of parameter combinations we verified dynamical states which turned out not to be reproduced by the reduced system. As discussed in Sec.~\ref{sec:accuracy}, we verified vanishing values for the temporal average of the mean-field frequency for couplings inscribed in a TW region. By visualizing the time-series of such unanticipated states, we found that one local parameter evolved with an oscillatory dynamics akin to breathing chimera states~\cite{abrams2008solvable}, in sharp contrast to the evolution predicted by our calculations for the same parameters and initial conditions. It is worth noting, nonetheless, that disagreements of this nature are somewhat expected to occur: for the identical frequencies case there exists a one-parameter family of invariant manifolds (of which the OA-manifold is a special solution) that are neutrally stable with respect to perturbations in directions transverse to themselves~\cite{pikovsky2008partially,pikovsky2015dynamics,martens2010bistable,marvel2009identical}. Hence, there could be certain types of perturbations that may drive the system away from the manifold contemplated by the OA ansatz, thus generating unexpected results such as the ones discussed in Sec.~\ref{sec:accuracy}. Another deviation from the theory was observed in the appearance of zero-lag sync and $\pi$-states over a region in the parameter space initially believed to manifest traveling-waves solely [see Fig.~\ref{fig4} (c)]. Future works should further investigate the emergence of chimeras and other states with respect to perturbations off the OA manifold for populations of identical oscillators coupled asymmetrically. 

\begin{figure*}[t!]
	\centering
	\includegraphics[width=2.0\columnwidth]{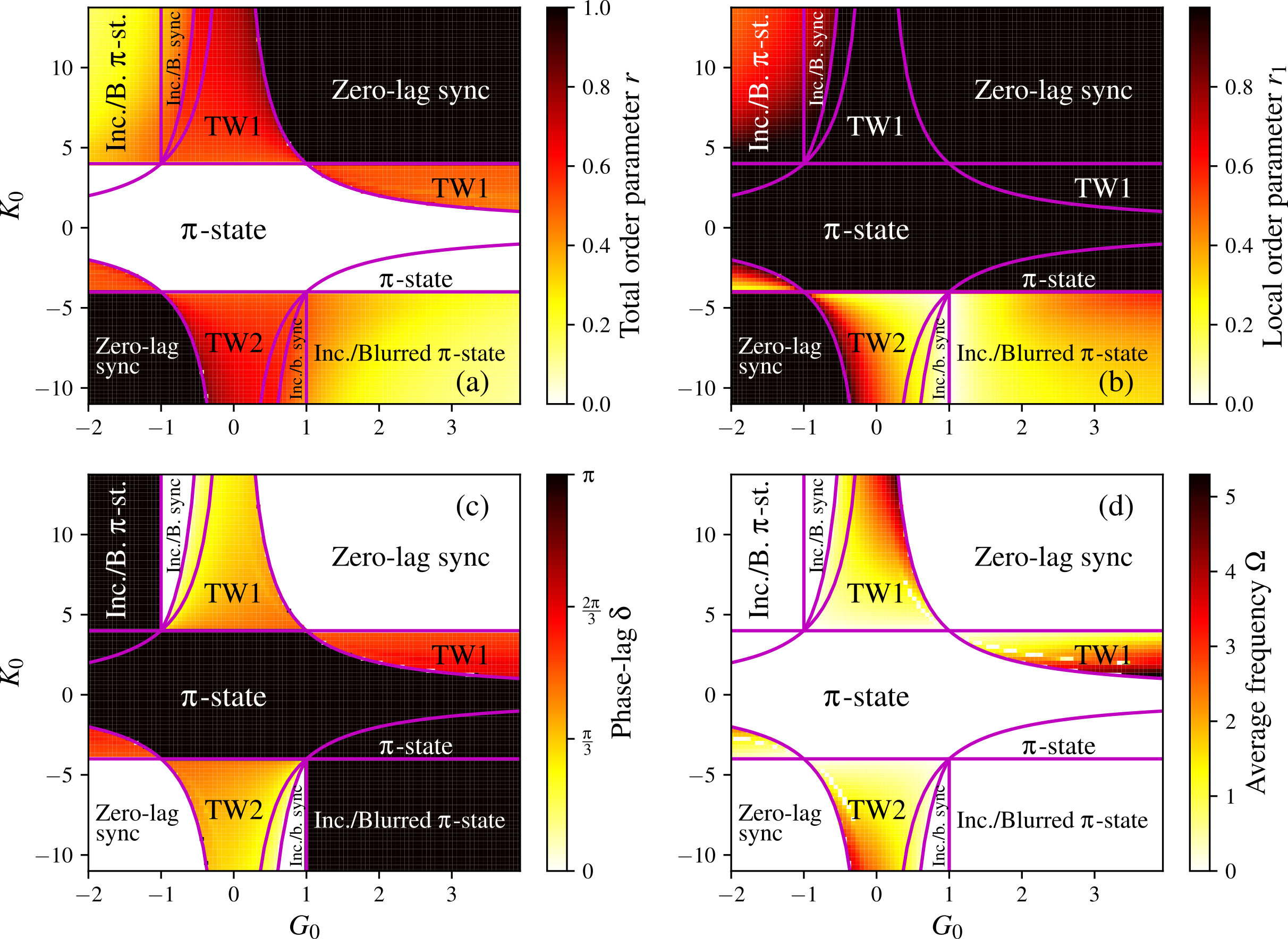}
	\caption{Comparison between simulations (colormaps) and theory (solid lines). (a) Total order parameter $r$; (b) local order parameter $r_1$ of subpopulation 1; (c) phase-lag $\delta$ measuring the separation between the two subpopulations; and (d) average frequency $\Omega$ [Eq.~(\ref{eq:avg_freq_OMEGA})]. For each pair of coupling $(G_0,K_0)$, Eqs.~(\ref{eq:KGmodel}) are  integrated numerically with the Heun's method considering $N=10^4$ oscillators and with integration time step $dt=0.005$. The long-time behavior of each parameter is quantified by averaging the trajectories over $t \in [500,1000]$. For all coupling pairs $(G_0,K_0)$, Eqs.~(\ref{eq:KGmodel}) were initiated with the exact same configuration for $\theta_i (t=0)$: phases in subpopulation 1 were randomly chosen from a Gaussian distribution with mean $\theta_1 \simeq 0.17 $ and standard deviation $\sigma_1 \simeq 0.85 $, yielding $r_1(t=0) \simeq 0.69$; phases of subpopulation 2 were also Gaussian distributed, but with mean $\theta_2 \simeq 1.09$ and standard deviation $\sigma_2 \simeq 1.053$, yielding $r_2(t=0) \simeq 0.57$. Coupling mismatch parameters: $\Delta K = 8$ and $\Delta G = 2$. $G_0 \times K_0 $ grid resolution: $100\times 100$ couplings.}
	\label{fig8}
\end{figure*}

As mentioned in the introduction, there are many systems whose dynamics can be modeled by phase oscillators interacting via positive and negative couplings. Our results can thereby serve as a guide in the search for clustered states in different contexts. For instance, as shown in Ref.~\cite{myung2015gaba}, phase-attraction or phase-repulsion alone cannot account for the regulation of circadian rhythms; a phase model incorporating mixed couplings linked asymmetrically, on the other hand, does reproduce the outcome of experiments with neuronal networks of the suprachiasmatic nucleus (SCN). Therefore, system (\ref{eq:KGmodel}) with two intertwined subpopulations may be a suitable model to describe the synchronization between the dorsal and ventral subregions of the SCN~\cite{myung2015gaba}. 

Finally, there are a number of potentially relevant extensions for the present model: given the non-trivial behavior uncovered here, it would be interesting, for instance, to investigate more than two coupled populations, as well as to 
study the effect of attractive and repulsive interactions on the collective dynamics of oscillators with higher-order harmonics in the coupling function~\cite{gong2019low,skardal2019abrupt}. One could also consider oscillators coupled on structures that allow interactions beyond the classical pairwise, such as hypergraphs~\cite{arruda2020social} and simplicial complexes~\cite{millan2020explosive}. On the experimental domain, realizations of the dynamical transitions reported here may be obtained in populations of chemical~\cite{totz2018spiral,cualuguaru2020first} and optical arrays~\cite{hagerstrom2012experimental}, which are  experimental setups that have been shown to reproduce chimeras and other dynamical states found in phase oscillator models.

\begin{acknowledgments}
The author thanks Bernard Sonnenschein, Chen Chris Gong, Deniz Eroglu, Paul Schultz, Vinicius Sciuti, and Bruno Messias for useful conversations. 
This research was funded by FAPESP (Grant No.~2016/23827-6) and carried out using the computational resources of the Center for Mathematical Sciences Applied to Industry (CeMEAI) funded by FAPESP (Grant No.~2013/07375-0).
\end{acknowledgments}

\appendix

\section{Supplementary diagram}

In this section we recalculate the diagrams of Fig.~\ref{fig5}
by initiating the oscillators differently than in Sec.~\ref{sec:simulations}. Specifically, in Fig.~\ref{fig8} we choose the phases of each subpopulation to be distributed according to distinct Gaussian distributions whose peaks are separated by a phase-lag (see the caption of Fig.~\ref{fig8} for details).  Comparing Fig.~\ref{fig5} with Fig.~\ref{fig8} we see that a blurred $\pi$-state region in the former is converted into a $\pi$-state area in the latter [notice the regions with $r_1=0$ in Fig.~\ref{fig5}(b) and $r_1 = 1$ in Fig.~\ref{fig8}(b)]. In addition to the bistability between TWs and blurred zero-lag sync states confirmed in Sec.~\ref{sec:accuracy}, another significant difference between Fig.~\ref{fig5} and Fig.~\ref{fig8} lies in the ``Incoherence/Blurred $\pi$-state'' areas: in Fig.~\ref{fig5} these regions exhibit small values for the order parameters (a consequence of choosing the initial conditions uniformly at random between $[-\pi,\pi]$), along with $\delta = \pi$; in Fig.~\ref{fig8}, on the other hand, we observe higher values for $r_1$, thus confirming that multiple local synchronization levels are possible in those regions as revealed by the analysis in Sec.~\ref{sec:bifurcation_analysis}.

%\begin{thebibliography}
\bibliography{bibliography}% Produces the bibliography via BibTeX.

\begin{thebibliography}{54}%
\makeatletter
\providecommand \@ifxundefined [1]{%
 \@ifx{#1\undefined}
}%
\providecommand \@ifnum [1]{%
 \ifnum #1\expandafter \@firstoftwo
 \else \expandafter \@secondoftwo
 \fi
}%
\providecommand \@ifx [1]{%
 \ifx #1\expandafter \@firstoftwo
 \else \expandafter \@secondoftwo
 \fi
}%
\providecommand \natexlab [1]{#1}%
\providecommand \enquote  [1]{``#1''}%
\providecommand \bibnamefont  [1]{#1}%
\providecommand \bibfnamefont [1]{#1}%
\providecommand \citenamefont [1]{#1}%
\providecommand \href@noop [0]{\@secondoftwo}%
\providecommand \href [0]{\begingroup \@sanitize@url \@href}%
\providecommand \@href[1]{\@@startlink{#1}\@@href}%
\providecommand \@@href[1]{\endgroup#1\@@endlink}%
\providecommand \@sanitize@url [0]{\catcode `\\12\catcode `\$12\catcode
  `\&12\catcode `\#12\catcode `\^12\catcode `\_12\catcode `\%12\relax}%
\providecommand \@@startlink[1]{}%
\providecommand \@@endlink[0]{}%
\providecommand \url  [0]{\begingroup\@sanitize@url \@url }%
\providecommand \@url [1]{\endgroup\@href {#1}{\urlprefix }}%
\providecommand \urlprefix  [0]{URL }%
\providecommand \Eprint [0]{\href }%
\providecommand \doibase [0]{http://dx.doi.org/}%
\providecommand \selectlanguage [0]{\@gobble}%
\providecommand \bibinfo  [0]{\@secondoftwo}%
\providecommand \bibfield  [0]{\@secondoftwo}%
\providecommand \translation [1]{[#1]}%
\providecommand \BibitemOpen [0]{}%
\providecommand \bibitemStop [0]{}%
\providecommand \bibitemNoStop [0]{.\EOS\space}%
\providecommand \EOS [0]{\spacefactor3000\relax}%
\providecommand \BibitemShut  [1]{\csname bibitem#1\endcsname}%
\let\auto@bib@innerbib\@empty
%</preamble>
\bibitem [{\citenamefont {Acebr{\'o}n}\ \emph {et~al.}(2005)\citenamefont
  {Acebr{\'o}n}, \citenamefont {Bonilla}, \citenamefont {Vicente},
  \citenamefont {Ritort},\ and\ \citenamefont {Spigler}}]{acebron2005kuramoto}%
  \BibitemOpen
  \bibfield  {author} {\bibinfo {author} {\bibfnamefont {Juan~A}\ \bibnamefont
  {Acebr{\'o}n}}, \bibinfo {author} {\bibfnamefont {Luis~L}\ \bibnamefont
  {Bonilla}}, \bibinfo {author} {\bibfnamefont {Conrad J~P{\'e}rez}\
  \bibnamefont {Vicente}}, \bibinfo {author} {\bibfnamefont {F{\'e}lix}\
  \bibnamefont {Ritort}}, \ and\ \bibinfo {author} {\bibfnamefont {Renato}\
  \bibnamefont {Spigler}},\ }\bibfield  {title} {\enquote {\bibinfo {title}
  {The {K}uramoto model: A simple paradigm for synchronization phenomena},}\
  }\href@noop {} {\bibfield  {journal} {\bibinfo  {journal} {{R}eviews of
  {M}odern {P}hysics}\ }\textbf {\bibinfo {volume} {77}},\ \bibinfo {pages}
  {137} (\bibinfo {year} {2005})}\BibitemShut {NoStop}%
\bibitem [{\citenamefont {Rodrigues}\ \emph {et~al.}(2016)\citenamefont
  {Rodrigues}, \citenamefont {Peron}, \citenamefont {Ji},\ and\ \citenamefont
  {Kurths}}]{rodrigues2016kuramoto}%
  \BibitemOpen
  \bibfield  {author} {\bibinfo {author} {\bibfnamefont {Francisco~A}\
  \bibnamefont {Rodrigues}}, \bibinfo {author} {\bibfnamefont {Thomas K~DM}\
  \bibnamefont {Peron}}, \bibinfo {author} {\bibfnamefont {Peng}\ \bibnamefont
  {Ji}}, \ and\ \bibinfo {author} {\bibfnamefont {J{\"u}rgen}\ \bibnamefont
  {Kurths}},\ }\bibfield  {title} {\enquote {\bibinfo {title} {The {K}uramoto
  model in complex networks},}\ }\href@noop {} {\bibfield  {journal} {\bibinfo
  {journal} {Physics Reports}\ }\textbf {\bibinfo {volume} {610}},\ \bibinfo
  {pages} {1--98} (\bibinfo {year} {2016})}\BibitemShut {NoStop}%
\bibitem [{\citenamefont {Arenas}\ \emph {et~al.}(2008)\citenamefont {Arenas},
  \citenamefont {D{\'\i}az-Guilera}, \citenamefont {Kurths}, \citenamefont
  {Moreno},\ and\ \citenamefont {Zhou}}]{arenas2008synchronization}%
  \BibitemOpen
  \bibfield  {author} {\bibinfo {author} {\bibfnamefont {Alex}\ \bibnamefont
  {Arenas}}, \bibinfo {author} {\bibfnamefont {Albert}\ \bibnamefont
  {D{\'\i}az-Guilera}}, \bibinfo {author} {\bibfnamefont {Jurgen}\ \bibnamefont
  {Kurths}}, \bibinfo {author} {\bibfnamefont {Yamir}\ \bibnamefont {Moreno}},
  \ and\ \bibinfo {author} {\bibfnamefont {Changsong}\ \bibnamefont {Zhou}},\
  }\bibfield  {title} {\enquote {\bibinfo {title} {Synchronization in complex
  networks},}\ }\href@noop {} {\bibfield  {journal} {\bibinfo  {journal}
  {Physics Reports}\ }\textbf {\bibinfo {volume} {469}},\ \bibinfo {pages}
  {93--153} (\bibinfo {year} {2008})}\BibitemShut {NoStop}%
\bibitem [{\citenamefont {Winfree}(1967)}]{winfree1967biological}%
  \BibitemOpen
  \bibfield  {author} {\bibinfo {author} {\bibfnamefont {Arthur~T}\
  \bibnamefont {Winfree}},\ }\bibfield  {title} {\enquote {\bibinfo {title}
  {Biological rhythms and the behavior of populations of coupled
  oscillators},}\ }\href@noop {} {\bibfield  {journal} {\bibinfo  {journal}
  {Journal of Theoretical Biology}\ }\textbf {\bibinfo {volume} {16}},\
  \bibinfo {pages} {15--42} (\bibinfo {year} {1967})}\BibitemShut {NoStop}%
\bibitem [{\citenamefont {Heinrich}\ \emph {et~al.}(2011)\citenamefont
  {Heinrich}, \citenamefont {Ludwig}, \citenamefont {Qian}, \citenamefont
  {Kubala},\ and\ \citenamefont {Marquardt}}]{heinrich2011collective}%
  \BibitemOpen
  \bibfield  {author} {\bibinfo {author} {\bibfnamefont {Georg}\ \bibnamefont
  {Heinrich}}, \bibinfo {author} {\bibfnamefont {Max}\ \bibnamefont {Ludwig}},
  \bibinfo {author} {\bibfnamefont {Jiang}\ \bibnamefont {Qian}}, \bibinfo
  {author} {\bibfnamefont {Bj{\"o}rn}\ \bibnamefont {Kubala}}, \ and\ \bibinfo
  {author} {\bibfnamefont {Florian}\ \bibnamefont {Marquardt}},\ }\bibfield
  {title} {\enquote {\bibinfo {title} {Collective dynamics in optomechanical
  arrays},}\ }\href@noop {} {\bibfield  {journal} {\bibinfo  {journal}
  {Physical {R}eview {L}etters}\ }\textbf {\bibinfo {volume} {107}},\ \bibinfo
  {pages} {043603} (\bibinfo {year} {2011})}\BibitemShut {NoStop}%
\bibitem [{\citenamefont {Wiesenfeld}\ \emph {et~al.}(1996)\citenamefont
  {Wiesenfeld}, \citenamefont {Colet},\ and\ \citenamefont
  {Strogatz}}]{wiesenfeld1996synchronization}%
  \BibitemOpen
  \bibfield  {author} {\bibinfo {author} {\bibfnamefont {Kurt}\ \bibnamefont
  {Wiesenfeld}}, \bibinfo {author} {\bibfnamefont {Pere}\ \bibnamefont
  {Colet}}, \ and\ \bibinfo {author} {\bibfnamefont {Steven~H}\ \bibnamefont
  {Strogatz}},\ }\bibfield  {title} {\enquote {\bibinfo {title}
  {Synchronization transitions in a disordered {J}osephson series array},}\
  }\href@noop {} {\bibfield  {journal} {\bibinfo  {journal} {Physical {R}eview
  {L}etters}\ }\textbf {\bibinfo {volume} {76}},\ \bibinfo {pages} {404}
  (\bibinfo {year} {1996})}\BibitemShut {NoStop}%
\bibitem [{\citenamefont {Kiss}\ \emph {et~al.}(2002)\citenamefont {Kiss},
  \citenamefont {Zhai},\ and\ \citenamefont {Hudson}}]{kiss2002emerging}%
  \BibitemOpen
  \bibfield  {author} {\bibinfo {author} {\bibfnamefont {Istv{\'a}n~Z}\
  \bibnamefont {Kiss}}, \bibinfo {author} {\bibfnamefont {Yumei}\ \bibnamefont
  {Zhai}}, \ and\ \bibinfo {author} {\bibfnamefont {John~L}\ \bibnamefont
  {Hudson}},\ }\bibfield  {title} {\enquote {\bibinfo {title} {Emerging
  coherence in a population of chemical oscillators},}\ }\href@noop {}
  {\bibfield  {journal} {\bibinfo  {journal} {Science}\ }\textbf {\bibinfo
  {volume} {296}},\ \bibinfo {pages} {1676--1678} (\bibinfo {year}
  {2002})}\BibitemShut {NoStop}%
\bibitem [{\citenamefont {D{\"o}rfler}\ and\ \citenamefont
  {Bullo}(2014)}]{dorfler2014synchronization}%
  \BibitemOpen
  \bibfield  {author} {\bibinfo {author} {\bibfnamefont {Florian}\ \bibnamefont
  {D{\"o}rfler}}\ and\ \bibinfo {author} {\bibfnamefont {Francesco}\
  \bibnamefont {Bullo}},\ }\bibfield  {title} {\enquote {\bibinfo {title}
  {Synchronization in complex networks of phase oscillators: A survey},}\
  }\href@noop {} {\bibfield  {journal} {\bibinfo  {journal} {Automatica}\
  }\textbf {\bibinfo {volume} {50}},\ \bibinfo {pages} {1539--1564} (\bibinfo
  {year} {2014})}\BibitemShut {NoStop}%
\bibitem [{\citenamefont {Shahal}\ \emph {et~al.}(2020)\citenamefont {Shahal},
  \citenamefont {Wurzberg}, \citenamefont {Sibony}, \citenamefont {Duadi},
  \citenamefont {Shniderman}, \citenamefont {Weymouth}, \citenamefont
  {Davidson},\ and\ \citenamefont {Fridman}}]{shahal2020synchronization}%
  \BibitemOpen
  \bibfield  {author} {\bibinfo {author} {\bibfnamefont {Shir}\ \bibnamefont
  {Shahal}}, \bibinfo {author} {\bibfnamefont {Ateret}\ \bibnamefont
  {Wurzberg}}, \bibinfo {author} {\bibfnamefont {Inbar}\ \bibnamefont
  {Sibony}}, \bibinfo {author} {\bibfnamefont {Hamootal}\ \bibnamefont
  {Duadi}}, \bibinfo {author} {\bibfnamefont {Elad}\ \bibnamefont
  {Shniderman}}, \bibinfo {author} {\bibfnamefont {Daniel}\ \bibnamefont
  {Weymouth}}, \bibinfo {author} {\bibfnamefont {Nir}\ \bibnamefont
  {Davidson}}, \ and\ \bibinfo {author} {\bibfnamefont {Moti}\ \bibnamefont
  {Fridman}},\ }\bibfield  {title} {\enquote {\bibinfo {title} {Synchronization
  of complex human networks},}\ }\href@noop {} {\bibfield  {journal} {\bibinfo
  {journal} {Nature Communications}\ }\textbf {\bibinfo {volume} {11}},\
  \bibinfo {pages} {1--10} (\bibinfo {year} {2020})}\BibitemShut {NoStop}%
\bibitem [{\citenamefont {Daido}(1992)}]{daido1992quasientrainment}%
  \BibitemOpen
  \bibfield  {author} {\bibinfo {author} {\bibfnamefont {Hiroaki}\ \bibnamefont
  {Daido}},\ }\bibfield  {title} {\enquote {\bibinfo {title} {Quasientrainment
  and slow relaxation in a population of oscillators with random and frustrated
  interactions},}\ }\href@noop {} {\bibfield  {journal} {\bibinfo  {journal}
  {Physical {R}eview {L}etters}\ }\textbf {\bibinfo {volume} {68}},\ \bibinfo
  {pages} {1073} (\bibinfo {year} {1992})}\BibitemShut {NoStop}%
\bibitem [{\citenamefont {Bonilla}\ \emph {et~al.}(1993)\citenamefont
  {Bonilla}, \citenamefont {Vicente},\ and\ \citenamefont
  {Rubi}}]{bonilla1993glassy}%
  \BibitemOpen
  \bibfield  {author} {\bibinfo {author} {\bibfnamefont {LL}~\bibnamefont
  {Bonilla}}, \bibinfo {author} {\bibfnamefont {CJ~P{\'e}rez}\ \bibnamefont
  {Vicente}}, \ and\ \bibinfo {author} {\bibfnamefont {JM}~\bibnamefont
  {Rubi}},\ }\bibfield  {title} {\enquote {\bibinfo {title} {Glassy
  synchronization in a population of coupled oscillators},}\ }\href@noop {}
  {\bibfield  {journal} {\bibinfo  {journal} {Journal of Statistical Physics}\
  }\textbf {\bibinfo {volume} {70}},\ \bibinfo {pages} {921--937} (\bibinfo
  {year} {1993})}\BibitemShut {NoStop}%
\bibitem [{\citenamefont {Stiller}\ and\ \citenamefont
  {Radons}(1998)}]{stiller1998dynamics}%
  \BibitemOpen
  \bibfield  {author} {\bibinfo {author} {\bibfnamefont {JC}~\bibnamefont
  {Stiller}}\ and\ \bibinfo {author} {\bibfnamefont {G}~\bibnamefont
  {Radons}},\ }\bibfield  {title} {\enquote {\bibinfo {title} {Dynamics of
  nonlinear oscillators with random interactions},}\ }\href@noop {} {\bibfield
  {journal} {\bibinfo  {journal} {Physical {R}eview E}\ }\textbf {\bibinfo
  {volume} {58}},\ \bibinfo {pages} {1789} (\bibinfo {year}
  {1998})}\BibitemShut {NoStop}%
\bibitem [{\citenamefont {Daido}(2000)}]{daido2000algebraic}%
  \BibitemOpen
  \bibfield  {author} {\bibinfo {author} {\bibfnamefont {Hiroaki}\ \bibnamefont
  {Daido}},\ }\bibfield  {title} {\enquote {\bibinfo {title} {Algebraic
  relaxation of an order parameter in randomly coupled limit-cycle
  oscillators},}\ }\href@noop {} {\bibfield  {journal} {\bibinfo  {journal}
  {Physical {R}eview E}\ }\textbf {\bibinfo {volume} {61}},\ \bibinfo {pages}
  {2145} (\bibinfo {year} {2000})}\BibitemShut {NoStop}%
\bibitem [{\citenamefont {Stiller}\ and\ \citenamefont
  {Radons}(2000)}]{stiller2000self}%
  \BibitemOpen
  \bibfield  {author} {\bibinfo {author} {\bibfnamefont {JC}~\bibnamefont
  {Stiller}}\ and\ \bibinfo {author} {\bibfnamefont {G}~\bibnamefont
  {Radons}},\ }\bibfield  {title} {\enquote {\bibinfo {title} {Self-averaging
  of an order parameter in randomly coupled limit-cycle oscillators},}\
  }\href@noop {} {\bibfield  {journal} {\bibinfo  {journal} {Physical {R}eview
  E}\ }\textbf {\bibinfo {volume} {61}},\ \bibinfo {pages} {2148} (\bibinfo
  {year} {2000})}\BibitemShut {NoStop}%
\bibitem [{\citenamefont {Iatsenko}\ \emph {et~al.}(2014)\citenamefont
  {Iatsenko}, \citenamefont {McClintock},\ and\ \citenamefont
  {Stefanovska}}]{iatsenko2014glassy}%
  \BibitemOpen
  \bibfield  {author} {\bibinfo {author} {\bibfnamefont {Dima}\ \bibnamefont
  {Iatsenko}}, \bibinfo {author} {\bibfnamefont {Peter~VE}\ \bibnamefont
  {McClintock}}, \ and\ \bibinfo {author} {\bibfnamefont {Aneta}\ \bibnamefont
  {Stefanovska}},\ }\bibfield  {title} {\enquote {\bibinfo {title} {Glassy
  states and super-relaxation in populations of coupled phase oscillators},}\
  }\href@noop {} {\bibfield  {journal} {\bibinfo  {journal} {Nature
  communications}\ }\textbf {\bibinfo {volume} {5}},\ \bibinfo {pages} {4118}
  (\bibinfo {year} {2014})}\BibitemShut {NoStop}%
\bibitem [{\citenamefont {Zanette}(2005)}]{zanette2005synchronization}%
  \BibitemOpen
  \bibfield  {author} {\bibinfo {author} {\bibfnamefont {Dami{\'a}n~H}\
  \bibnamefont {Zanette}},\ }\bibfield  {title} {\enquote {\bibinfo {title}
  {Synchronization and frustration in oscillator networks with attractive and
  repulsive interactions},}\ }\href@noop {} {\bibfield  {journal} {\bibinfo
  {journal} {EPL (Europhysics {L}etters)}\ }\textbf {\bibinfo {volume} {72}},\
  \bibinfo {pages} {190} (\bibinfo {year} {2005})}\BibitemShut {NoStop}%
\bibitem [{\citenamefont {Hong}\ and\ \citenamefont
  {Strogatz}(2011{\natexlab{a}})}]{hong2011kuramoto}%
  \BibitemOpen
  \bibfield  {author} {\bibinfo {author} {\bibfnamefont {Hyunsuk}\ \bibnamefont
  {Hong}}\ and\ \bibinfo {author} {\bibfnamefont {Steven~H}\ \bibnamefont
  {Strogatz}},\ }\bibfield  {title} {\enquote {\bibinfo {title} {{K}uramoto
  model of coupled oscillators with positive and negative coupling parameters:
  an example of conformist and contrarian oscillators},}\ }\href@noop {}
  {\bibfield  {journal} {\bibinfo  {journal} {Physical {R}eview {L}etters}\
  }\textbf {\bibinfo {volume} {106}},\ \bibinfo {pages} {054102} (\bibinfo
  {year} {2011}{\natexlab{a}})}\BibitemShut {NoStop}%
\bibitem [{\citenamefont {Hong}\ and\ \citenamefont
  {Strogatz}(2011{\natexlab{b}})}]{hong2011conformists}%
  \BibitemOpen
  \bibfield  {author} {\bibinfo {author} {\bibfnamefont {Hyunsuk}\ \bibnamefont
  {Hong}}\ and\ \bibinfo {author} {\bibfnamefont {Steven~H}\ \bibnamefont
  {Strogatz}},\ }\bibfield  {title} {\enquote {\bibinfo {title} {Conformists
  and contrarians in a {K}uramoto model with identical natural frequencies},}\
  }\href@noop {} {\bibfield  {journal} {\bibinfo  {journal} {Physical {R}eview
  E}\ }\textbf {\bibinfo {volume} {84}},\ \bibinfo {pages} {046202} (\bibinfo
  {year} {2011}{\natexlab{b}})}\BibitemShut {NoStop}%
\bibitem [{\citenamefont {Hong}\ and\ \citenamefont
  {Strogatz}(2012)}]{hong2012mean}%
  \BibitemOpen
  \bibfield  {author} {\bibinfo {author} {\bibfnamefont {Hyunsuk}\ \bibnamefont
  {Hong}}\ and\ \bibinfo {author} {\bibfnamefont {Steven~H}\ \bibnamefont
  {Strogatz}},\ }\bibfield  {title} {\enquote {\bibinfo {title} {Mean-field
  behavior in coupled oscillators with attractive and repulsive
  interactions},}\ }\href@noop {} {\bibfield  {journal} {\bibinfo  {journal}
  {Physical {R}eview E}\ }\textbf {\bibinfo {volume} {85}},\ \bibinfo {pages}
  {056210} (\bibinfo {year} {2012})}\BibitemShut {NoStop}%
\bibitem [{\citenamefont {Montbri{\'o}}\ and\ \citenamefont
  {Paz{\'o}}(2011)}]{montbrio2011collective}%
  \BibitemOpen
  \bibfield  {author} {\bibinfo {author} {\bibfnamefont {Ernest}\ \bibnamefont
  {Montbri{\'o}}}\ and\ \bibinfo {author} {\bibfnamefont {Diego}\ \bibnamefont
  {Paz{\'o}}},\ }\bibfield  {title} {\enquote {\bibinfo {title} {Collective
  synchronization in the presence of reactive coupling and shear diversity},}\
  }\href@noop {} {\bibfield  {journal} {\bibinfo  {journal} {Physical {R}eview
  E}\ }\textbf {\bibinfo {volume} {84}},\ \bibinfo {pages} {046206} (\bibinfo
  {year} {2011})}\BibitemShut {NoStop}%
\bibitem [{\citenamefont {Iatsenko}\ \emph {et~al.}(2013)\citenamefont
  {Iatsenko}, \citenamefont {Petkoski}, \citenamefont {McClintock},\ and\
  \citenamefont {Stefanovska}}]{iatsenko2013stationary}%
  \BibitemOpen
  \bibfield  {author} {\bibinfo {author} {\bibfnamefont {Dmytro}\ \bibnamefont
  {Iatsenko}}, \bibinfo {author} {\bibfnamefont {Spase}\ \bibnamefont
  {Petkoski}}, \bibinfo {author} {\bibfnamefont {PVE}\ \bibnamefont
  {McClintock}}, \ and\ \bibinfo {author} {\bibfnamefont {A}~\bibnamefont
  {Stefanovska}},\ }\bibfield  {title} {\enquote {\bibinfo {title} {Stationary
  and traveling wave states of the {K}uramoto model with an arbitrary
  distribution of frequencies and coupling strengths},}\ }\href@noop {}
  {\bibfield  {journal} {\bibinfo  {journal} {Physical {R}eview {L}etters}\
  }\textbf {\bibinfo {volume} {110}},\ \bibinfo {pages} {064101} (\bibinfo
  {year} {2013})}\BibitemShut {NoStop}%
\bibitem [{\citenamefont {Hong}(2014)}]{hong2014periodic}%
  \BibitemOpen
  \bibfield  {author} {\bibinfo {author} {\bibfnamefont {Hyunsuk}\ \bibnamefont
  {Hong}},\ }\bibfield  {title} {\enquote {\bibinfo {title} {Periodic
  synchronization and chimera in conformist and contrarian oscillators},}\
  }\href@noop {} {\bibfield  {journal} {\bibinfo  {journal} {Physical {R}eview
  E}\ }\textbf {\bibinfo {volume} {89}},\ \bibinfo {pages} {062924} (\bibinfo
  {year} {2014})}\BibitemShut {NoStop}%
\bibitem [{\citenamefont {Kloumann}\ \emph {et~al.}(2014)\citenamefont
  {Kloumann}, \citenamefont {Lizarraga},\ and\ \citenamefont
  {Strogatz}}]{kloumann2014phase}%
  \BibitemOpen
  \bibfield  {author} {\bibinfo {author} {\bibfnamefont {Isabel~M}\
  \bibnamefont {Kloumann}}, \bibinfo {author} {\bibfnamefont {Ian~M}\
  \bibnamefont {Lizarraga}}, \ and\ \bibinfo {author} {\bibfnamefont
  {Steven~H}\ \bibnamefont {Strogatz}},\ }\bibfield  {title} {\enquote
  {\bibinfo {title} {Phase diagram for the {K}uramoto model with van hemmen
  interactions},}\ }\href@noop {} {\bibfield  {journal} {\bibinfo  {journal}
  {Physical Review E}\ }\textbf {\bibinfo {volume} {89}},\ \bibinfo {pages}
  {012904} (\bibinfo {year} {2014})}\BibitemShut {NoStop}%
\bibitem [{\citenamefont {Sonnenschein}\ \emph {et~al.}(2015)\citenamefont
  {Sonnenschein}, \citenamefont {Peron}, \citenamefont {Rodrigues},
  \citenamefont {Kurths},\ and\ \citenamefont
  {Schimansky-Geier}}]{sonnenschein2015collective}%
  \BibitemOpen
  \bibfield  {author} {\bibinfo {author} {\bibfnamefont {Bernard}\ \bibnamefont
  {Sonnenschein}}, \bibinfo {author} {\bibfnamefont {Thomas K~DM}\ \bibnamefont
  {Peron}}, \bibinfo {author} {\bibfnamefont {Francisco~A}\ \bibnamefont
  {Rodrigues}}, \bibinfo {author} {\bibfnamefont {J{\"u}rgen}\ \bibnamefont
  {Kurths}}, \ and\ \bibinfo {author} {\bibfnamefont {Lutz}\ \bibnamefont
  {Schimansky-Geier}},\ }\bibfield  {title} {\enquote {\bibinfo {title}
  {Collective dynamics in two populations of noisy oscillators with asymmetric
  interactions},}\ }\href@noop {} {\bibfield  {journal} {\bibinfo  {journal}
  {Physical {R}eview E}\ }\textbf {\bibinfo {volume} {91}},\ \bibinfo {pages}
  {062910} (\bibinfo {year} {2015})}\BibitemShut {NoStop}%
\bibitem [{\citenamefont {Ottino-L{\"o}ffler}\ and\ \citenamefont
  {Strogatz}(2018)}]{ottino2018volcano}%
  \BibitemOpen
  \bibfield  {author} {\bibinfo {author} {\bibfnamefont {Bertrand}\
  \bibnamefont {Ottino-L{\"o}ffler}}\ and\ \bibinfo {author} {\bibfnamefont
  {Steven~H}\ \bibnamefont {Strogatz}},\ }\bibfield  {title} {\enquote
  {\bibinfo {title} {Volcano transition in a solvable model of frustrated
  oscillators},}\ }\href@noop {} {\bibfield  {journal} {\bibinfo  {journal}
  {Physical {R}eview {L}etters}\ }\textbf {\bibinfo {volume} {120}},\ \bibinfo
  {pages} {264102} (\bibinfo {year} {2018})}\BibitemShut {NoStop}%
\bibitem [{\citenamefont {Park}\ and\ \citenamefont
  {Kahng}(2018)}]{park2018metastable}%
  \BibitemOpen
  \bibfield  {author} {\bibinfo {author} {\bibfnamefont {Jinha}\ \bibnamefont
  {Park}}\ and\ \bibinfo {author} {\bibfnamefont {B}~\bibnamefont {Kahng}},\
  }\bibfield  {title} {\enquote {\bibinfo {title} {Metastable state en route to
  traveling-wave synchronization state},}\ }\href@noop {} {\bibfield  {journal}
  {\bibinfo  {journal} {Physical {R}eview E}\ }\textbf {\bibinfo {volume}
  {97}},\ \bibinfo {pages} {020203} (\bibinfo {year} {2018})}\BibitemShut
  {NoStop}%
\bibitem [{\citenamefont {Anderson}\ \emph {et~al.}(2012)\citenamefont
  {Anderson}, \citenamefont {Tenzer}, \citenamefont {Barlev}, \citenamefont
  {Girvan}, \citenamefont {Antonsen},\ and\ \citenamefont
  {Ott}}]{anderson2012multiscale}%
  \BibitemOpen
  \bibfield  {author} {\bibinfo {author} {\bibfnamefont {Dustin}\ \bibnamefont
  {Anderson}}, \bibinfo {author} {\bibfnamefont {Ari}\ \bibnamefont {Tenzer}},
  \bibinfo {author} {\bibfnamefont {Gilad}\ \bibnamefont {Barlev}}, \bibinfo
  {author} {\bibfnamefont {Michelle}\ \bibnamefont {Girvan}}, \bibinfo {author}
  {\bibfnamefont {Thomas~M}\ \bibnamefont {Antonsen}}, \ and\ \bibinfo {author}
  {\bibfnamefont {Edward}\ \bibnamefont {Ott}},\ }\bibfield  {title} {\enquote
  {\bibinfo {title} {Multiscale dynamics in communities of phase
  oscillators},}\ }\href@noop {} {\bibfield  {journal} {\bibinfo  {journal}
  {Chaos: An Interdisciplinary Journal of Nonlinear Science}\ }\textbf
  {\bibinfo {volume} {22}},\ \bibinfo {pages} {013102} (\bibinfo {year}
  {2012})}\BibitemShut {NoStop}%
\bibitem [{\citenamefont {Park}\ and\ \citenamefont
  {Kahng}(2020)}]{park2020competing}%
  \BibitemOpen
  \bibfield  {author} {\bibinfo {author} {\bibfnamefont {Jinha}\ \bibnamefont
  {Park}}\ and\ \bibinfo {author} {\bibfnamefont {B}~\bibnamefont {Kahng}},\
  }\bibfield  {title} {\enquote {\bibinfo {title} {Competing synchronization on
  random networks},}\ }\href@noop {} {\bibfield  {journal} {\bibinfo  {journal}
  {Journal of Statistical Mechanics: Theory and Experiment}\ }\textbf {\bibinfo
  {volume} {2020}},\ \bibinfo {pages} {073407} (\bibinfo {year}
  {2020})}\BibitemShut {NoStop}%
\bibitem [{\citenamefont {Sonnenschein}\ and\ \citenamefont
  {Schimansky-Geier}(2013)}]{sonnenschein2013approximate}%
  \BibitemOpen
  \bibfield  {author} {\bibinfo {author} {\bibfnamefont {Bernard}\ \bibnamefont
  {Sonnenschein}}\ and\ \bibinfo {author} {\bibfnamefont {Lutz}\ \bibnamefont
  {Schimansky-Geier}},\ }\bibfield  {title} {\enquote {\bibinfo {title}
  {Approximate solution to the stochastic {K}uramoto model},}\ }\href@noop {}
  {\bibfield  {journal} {\bibinfo  {journal} {Physical {R}eview E}\ }\textbf
  {\bibinfo {volume} {88}},\ \bibinfo {pages} {052111} (\bibinfo {year}
  {2013})}\BibitemShut {NoStop}%
\bibitem [{\citenamefont {Tradonsky}\ \emph {et~al.}(2015)\citenamefont
  {Tradonsky}, \citenamefont {Nixon}, \citenamefont {Ronen}, \citenamefont
  {Pal}, \citenamefont {Chriki}, \citenamefont {Friesem},\ and\ \citenamefont
  {Davidson}}]{tradonsky2015conversion}%
  \BibitemOpen
  \bibfield  {author} {\bibinfo {author} {\bibfnamefont {Chene}\ \bibnamefont
  {Tradonsky}}, \bibinfo {author} {\bibfnamefont {Micha}\ \bibnamefont
  {Nixon}}, \bibinfo {author} {\bibfnamefont {Eitan}\ \bibnamefont {Ronen}},
  \bibinfo {author} {\bibfnamefont {Vishwa}\ \bibnamefont {Pal}}, \bibinfo
  {author} {\bibfnamefont {Ronen}\ \bibnamefont {Chriki}}, \bibinfo {author}
  {\bibfnamefont {Asher~A}\ \bibnamefont {Friesem}}, \ and\ \bibinfo {author}
  {\bibfnamefont {Nir}\ \bibnamefont {Davidson}},\ }\bibfield  {title}
  {\enquote {\bibinfo {title} {Conversion of out-of-phase to in-phase order in
  coupled laser arrays with second harmonics},}\ }\href@noop {} {\bibfield
  {journal} {\bibinfo  {journal} {Photonics Research}\ }\textbf {\bibinfo
  {volume} {3}},\ \bibinfo {pages} {77--81} (\bibinfo {year}
  {2015})}\BibitemShut {NoStop}%
\bibitem [{\citenamefont {Pal}\ \emph {et~al.}(2020)\citenamefont {Pal},
  \citenamefont {Mahler}, \citenamefont {Tradonsky}, \citenamefont {Friesem},\
  and\ \citenamefont {Davidson}}]{pal2020rapid}%
  \BibitemOpen
  \bibfield  {author} {\bibinfo {author} {\bibfnamefont {Vishwa}\ \bibnamefont
  {Pal}}, \bibinfo {author} {\bibfnamefont {Simon}\ \bibnamefont {Mahler}},
  \bibinfo {author} {\bibfnamefont {Chene}\ \bibnamefont {Tradonsky}}, \bibinfo
  {author} {\bibfnamefont {Asher~A}\ \bibnamefont {Friesem}}, \ and\ \bibinfo
  {author} {\bibfnamefont {Nir}\ \bibnamefont {Davidson}},\ }\bibfield  {title}
  {\enquote {\bibinfo {title} {Rapid fair sampling of the x y spin hamiltonian
  with a laser simulator},}\ }\href@noop {} {\bibfield  {journal} {\bibinfo
  {journal} {Physical Review Research}\ }\textbf {\bibinfo {volume} {2}},\
  \bibinfo {pages} {033008} (\bibinfo {year} {2020})}\BibitemShut {NoStop}%
\bibitem [{\citenamefont {Kori}\ \emph {et~al.}(2018)\citenamefont {Kori},
  \citenamefont {Kiss}, \citenamefont {Jain},\ and\ \citenamefont
  {Hudson}}]{kori2018partial}%
  \BibitemOpen
  \bibfield  {author} {\bibinfo {author} {\bibfnamefont {Hiroshi}\ \bibnamefont
  {Kori}}, \bibinfo {author} {\bibfnamefont {Istv{\'a}n~Z}\ \bibnamefont
  {Kiss}}, \bibinfo {author} {\bibfnamefont {Swati}\ \bibnamefont {Jain}}, \
  and\ \bibinfo {author} {\bibfnamefont {John~L}\ \bibnamefont {Hudson}},\
  }\bibfield  {title} {\enquote {\bibinfo {title} {Partial synchronization of
  relaxation oscillators with repulsive coupling in autocatalytic
  integrate-and-fire model and electrochemical experiments},}\ }\href@noop {}
  {\bibfield  {journal} {\bibinfo  {journal} {Chaos: An Interdisciplinary
  Journal of Nonlinear Science}\ }\textbf {\bibinfo {volume} {28}},\ \bibinfo
  {pages} {045111} (\bibinfo {year} {2018})}\BibitemShut {NoStop}%
\bibitem [{\citenamefont {Sebek}\ and\ \citenamefont
  {Kiss}(2019)}]{sebek2019plasticity}%
  \BibitemOpen
  \bibfield  {author} {\bibinfo {author} {\bibfnamefont {Michael}\ \bibnamefont
  {Sebek}}\ and\ \bibinfo {author} {\bibfnamefont {Istv{\'a}n~Z}\ \bibnamefont
  {Kiss}},\ }\bibfield  {title} {\enquote {\bibinfo {title} {Plasticity
  facilitates pattern selection of networks of chemical oscillations},}\
  }\href@noop {} {\bibfield  {journal} {\bibinfo  {journal} {Chaos: An
  Interdisciplinary Journal of Nonlinear Science}\ }\textbf {\bibinfo {volume}
  {29}},\ \bibinfo {pages} {083117} (\bibinfo {year} {2019})}\BibitemShut
  {NoStop}%
\bibitem [{\citenamefont {Myung}\ \emph {et~al.}(2015)\citenamefont {Myung},
  \citenamefont {Hong}, \citenamefont {DeWoskin}, \citenamefont {De~Schutter},
  \citenamefont {Forger},\ and\ \citenamefont {Takumi}}]{myung2015gaba}%
  \BibitemOpen
  \bibfield  {author} {\bibinfo {author} {\bibfnamefont {Jihwan}\ \bibnamefont
  {Myung}}, \bibinfo {author} {\bibfnamefont {Sungho}\ \bibnamefont {Hong}},
  \bibinfo {author} {\bibfnamefont {Daniel}\ \bibnamefont {DeWoskin}}, \bibinfo
  {author} {\bibfnamefont {Erik}\ \bibnamefont {De~Schutter}}, \bibinfo
  {author} {\bibfnamefont {Daniel~B}\ \bibnamefont {Forger}}, \ and\ \bibinfo
  {author} {\bibfnamefont {Toru}\ \bibnamefont {Takumi}},\ }\bibfield  {title}
  {\enquote {\bibinfo {title} {Gaba-mediated repulsive coupling between
  circadian clock neurons in the scn encodes seasonal time},}\ }\href@noop {}
  {\bibfield  {journal} {\bibinfo  {journal} {Proceedings of the National
  Academy of Sciences}\ }\textbf {\bibinfo {volume} {112}},\ \bibinfo {pages}
  {E3920--E3929} (\bibinfo {year} {2015})}\BibitemShut {NoStop}%
\bibitem [{\citenamefont {Ott}\ and\ \citenamefont
  {Antonsen}(2008)}]{ott2008low}%
  \BibitemOpen
  \bibfield  {author} {\bibinfo {author} {\bibfnamefont {Edward}\ \bibnamefont
  {Ott}}\ and\ \bibinfo {author} {\bibfnamefont {Thomas~M}\ \bibnamefont
  {Antonsen}},\ }\bibfield  {title} {\enquote {\bibinfo {title} {Low
  dimensional behavior of large systems of globally coupled oscillators},}\
  }\href@noop {} {\bibfield  {journal} {\bibinfo  {journal} {Chaos: An
  Interdisciplinary Journal of Nonlinear Science}\ }\textbf {\bibinfo {volume}
  {18}},\ \bibinfo {pages} {037113} (\bibinfo {year} {2008})}\BibitemShut
  {NoStop}%
\bibitem [{\citenamefont {Pikovsky}\ and\ \citenamefont
  {Rosenblum}(2015)}]{pikovsky2015dynamics}%
  \BibitemOpen
  \bibfield  {author} {\bibinfo {author} {\bibfnamefont {Arkady}\ \bibnamefont
  {Pikovsky}}\ and\ \bibinfo {author} {\bibfnamefont {Michael}\ \bibnamefont
  {Rosenblum}},\ }\bibfield  {title} {\enquote {\bibinfo {title} {Dynamics of
  globally coupled oscillators: Progress and perspectives},}\ }\href@noop {}
  {\bibfield  {journal} {\bibinfo  {journal} {Chaos: An Interdisciplinary
  Journal of Nonlinear Science}\ }\textbf {\bibinfo {volume} {25}},\ \bibinfo
  {pages} {097616} (\bibinfo {year} {2015})}\BibitemShut {NoStop}%
\bibitem [{\citenamefont {Meylahn}(2020)}]{meylahn2020two}%
  \BibitemOpen
  \bibfield  {author} {\bibinfo {author} {\bibfnamefont {JM}~\bibnamefont
  {Meylahn}},\ }\bibfield  {title} {\enquote {\bibinfo {title} {Two-community
  noisy {K}uramoto model},}\ }\href@noop {} {\bibfield  {journal} {\bibinfo
  {journal} {Nonlinearity}\ }\textbf {\bibinfo {volume} {33}},\ \bibinfo
  {pages} {1847} (\bibinfo {year} {2020})}\BibitemShut {NoStop}%
\bibitem [{\citenamefont {Vlasov}\ \emph {et~al.}(2014)\citenamefont {Vlasov},
  \citenamefont {Macau},\ and\ \citenamefont
  {Pikovsky}}]{vlasov2014synchronization}%
  \BibitemOpen
  \bibfield  {author} {\bibinfo {author} {\bibfnamefont {Vladimir}\
  \bibnamefont {Vlasov}}, \bibinfo {author} {\bibfnamefont {Elbert~EN}\
  \bibnamefont {Macau}}, \ and\ \bibinfo {author} {\bibfnamefont {Arkady}\
  \bibnamefont {Pikovsky}},\ }\bibfield  {title} {\enquote {\bibinfo {title}
  {Synchronization of oscillators in a {K}uramoto-type model with generic
  coupling},}\ }\href@noop {} {\bibfield  {journal} {\bibinfo  {journal}
  {Chaos: An Interdisciplinary Journal of Nonlinear Science}\ }\textbf
  {\bibinfo {volume} {24}},\ \bibinfo {pages} {023120} (\bibinfo {year}
  {2014})}\BibitemShut {NoStop}%
\bibitem [{\citenamefont {Sakaguchi}\ and\ \citenamefont
  {{K}uramoto}(1986)}]{sakaguchi1986soluble}%
  \BibitemOpen
  \bibfield  {author} {\bibinfo {author} {\bibfnamefont {Hidetsugu}\
  \bibnamefont {Sakaguchi}}\ and\ \bibinfo {author} {\bibfnamefont {Yoshiki}\
  \bibnamefont {{K}uramoto}},\ }\bibfield  {title} {\enquote {\bibinfo {title}
  {A soluble active rotater model showing phase transitions via mutual
  entertainment},}\ }\href@noop {} {\bibfield  {journal} {\bibinfo  {journal}
  {Progress of Theoretical Physics}\ }\textbf {\bibinfo {volume} {76}},\
  \bibinfo {pages} {576--581} (\bibinfo {year} {1986})}\BibitemShut {NoStop}%
\bibitem [{\citenamefont {Peron}\ \emph {et~al.}(2016)\citenamefont {Peron},
  \citenamefont {Kurths}, \citenamefont {Rodrigues}, \citenamefont
  {Schimansky-Geier},\ and\ \citenamefont {Sonnenschein}}]{peron2016traveling}%
  \BibitemOpen
  \bibfield  {author} {\bibinfo {author} {\bibfnamefont {Thomas K~DM}\
  \bibnamefont {Peron}}, \bibinfo {author} {\bibfnamefont {J{\"u}rgen}\
  \bibnamefont {Kurths}}, \bibinfo {author} {\bibfnamefont {Francisco~A}\
  \bibnamefont {Rodrigues}}, \bibinfo {author} {\bibfnamefont {Lutz}\
  \bibnamefont {Schimansky-Geier}}, \ and\ \bibinfo {author} {\bibfnamefont
  {Bernard}\ \bibnamefont {Sonnenschein}},\ }\bibfield  {title} {\enquote
  {\bibinfo {title} {Traveling phase waves in asymmetric networks of noisy
  chaotic attractors},}\ }\href@noop {} {\bibfield  {journal} {\bibinfo
  {journal} {Physical Review E}\ }\textbf {\bibinfo {volume} {94}},\ \bibinfo
  {pages} {042210} (\bibinfo {year} {2016})}\BibitemShut {NoStop}%
\bibitem [{\citenamefont {Watanabe}\ and\ \citenamefont
  {Strogatz}(1994)}]{watanabe1994constants}%
  \BibitemOpen
  \bibfield  {author} {\bibinfo {author} {\bibfnamefont {Shinya}\ \bibnamefont
  {Watanabe}}\ and\ \bibinfo {author} {\bibfnamefont {Steven~H}\ \bibnamefont
  {Strogatz}},\ }\bibfield  {title} {\enquote {\bibinfo {title} {Constants of
  motion for superconducting josephson arrays},}\ }\href@noop {} {\bibfield
  {journal} {\bibinfo  {journal} {Physica D: Nonlinear Phenomena}\ }\textbf
  {\bibinfo {volume} {74}},\ \bibinfo {pages} {197--253} (\bibinfo {year}
  {1994})}\BibitemShut {NoStop}%
\bibitem [{\citenamefont {Pikovsky}\ and\ \citenamefont
  {Rosenblum}(2008)}]{pikovsky2008partially}%
  \BibitemOpen
  \bibfield  {author} {\bibinfo {author} {\bibfnamefont {Arkady}\ \bibnamefont
  {Pikovsky}}\ and\ \bibinfo {author} {\bibfnamefont {Michael}\ \bibnamefont
  {Rosenblum}},\ }\bibfield  {title} {\enquote {\bibinfo {title} {Partially
  integrable dynamics of hierarchical populations of coupled oscillators},}\
  }\href@noop {} {\bibfield  {journal} {\bibinfo  {journal} {Physical {R}eview
  {L}etters}\ }\textbf {\bibinfo {volume} {101}},\ \bibinfo {pages} {264103}
  (\bibinfo {year} {2008})}\BibitemShut {NoStop}%
\bibitem [{\citenamefont {Petkoski}\ \emph {et~al.}(2013)\citenamefont
  {Petkoski}, \citenamefont {Iatsenko}, \citenamefont {Basnarkov},\ and\
  \citenamefont {Stefanovska}}]{petkoski2013mean}%
  \BibitemOpen
  \bibfield  {author} {\bibinfo {author} {\bibfnamefont {Spase}\ \bibnamefont
  {Petkoski}}, \bibinfo {author} {\bibfnamefont {Dmytro}\ \bibnamefont
  {Iatsenko}}, \bibinfo {author} {\bibfnamefont {Lasko}\ \bibnamefont
  {Basnarkov}}, \ and\ \bibinfo {author} {\bibfnamefont {Aneta}\ \bibnamefont
  {Stefanovska}},\ }\bibfield  {title} {\enquote {\bibinfo {title} {Mean-field
  and mean-ensemble frequencies of a system of coupled oscillators},}\
  }\href@noop {} {\bibfield  {journal} {\bibinfo  {journal} {Physical {R}eview
  E}\ }\textbf {\bibinfo {volume} {87}},\ \bibinfo {pages} {032908} (\bibinfo
  {year} {2013})}\BibitemShut {NoStop}%
\bibitem [{\citenamefont {Abrams}\ \emph {et~al.}(2008)\citenamefont {Abrams},
  \citenamefont {Mirollo}, \citenamefont {Strogatz},\ and\ \citenamefont
  {Wiley}}]{abrams2008solvable}%
  \BibitemOpen
  \bibfield  {author} {\bibinfo {author} {\bibfnamefont {Daniel~M}\
  \bibnamefont {Abrams}}, \bibinfo {author} {\bibfnamefont {Rennie}\
  \bibnamefont {Mirollo}}, \bibinfo {author} {\bibfnamefont {Steven~H}\
  \bibnamefont {Strogatz}}, \ and\ \bibinfo {author} {\bibfnamefont {Daniel~A}\
  \bibnamefont {Wiley}},\ }\bibfield  {title} {\enquote {\bibinfo {title}
  {Solvable model for chimera states of coupled oscillators},}\ }\href@noop {}
  {\bibfield  {journal} {\bibinfo  {journal} {Physical {R}eview {L}etters}\
  }\textbf {\bibinfo {volume} {101}},\ \bibinfo {pages} {084103} (\bibinfo
  {year} {2008})}\BibitemShut {NoStop}%
\bibitem [{\citenamefont {Pietras}\ \emph {et~al.}(2016)\citenamefont
  {Pietras}, \citenamefont {Deschle},\ and\ \citenamefont
  {Daffertshofer}}]{pietras2016equivalence}%
  \BibitemOpen
  \bibfield  {author} {\bibinfo {author} {\bibfnamefont {Bastian}\ \bibnamefont
  {Pietras}}, \bibinfo {author} {\bibfnamefont {Nicol{\'a}s}\ \bibnamefont
  {Deschle}}, \ and\ \bibinfo {author} {\bibfnamefont {Andreas}\ \bibnamefont
  {Daffertshofer}},\ }\bibfield  {title} {\enquote {\bibinfo {title}
  {Equivalence of coupled networks and networks with multimodal frequency
  distributions: Conditions for the bimodal and trimodal case},}\ }\href@noop
  {} {\bibfield  {journal} {\bibinfo  {journal} {Physical {R}eview E}\ }\textbf
  {\bibinfo {volume} {94}},\ \bibinfo {pages} {052211} (\bibinfo {year}
  {2016})}\BibitemShut {NoStop}%
\bibitem [{\citenamefont {Martens}(2010)}]{martens2010bistable}%
  \BibitemOpen
  \bibfield  {author} {\bibinfo {author} {\bibfnamefont {Erik~A}\ \bibnamefont
  {Martens}},\ }\bibfield  {title} {\enquote {\bibinfo {title} {Bistable
  chimera attractors on a triangular network of oscillator populations},}\
  }\href@noop {} {\bibfield  {journal} {\bibinfo  {journal} {Physical Review
  E}\ }\textbf {\bibinfo {volume} {82}},\ \bibinfo {pages} {016216} (\bibinfo
  {year} {2010})}\BibitemShut {NoStop}%
\bibitem [{\citenamefont {Marvel}\ \emph {et~al.}(2009)\citenamefont {Marvel},
  \citenamefont {Mirollo},\ and\ \citenamefont
  {Strogatz}}]{marvel2009identical}%
  \BibitemOpen
  \bibfield  {author} {\bibinfo {author} {\bibfnamefont {Seth~A}\ \bibnamefont
  {Marvel}}, \bibinfo {author} {\bibfnamefont {Renato~E}\ \bibnamefont
  {Mirollo}}, \ and\ \bibinfo {author} {\bibfnamefont {Steven~H}\ \bibnamefont
  {Strogatz}},\ }\bibfield  {title} {\enquote {\bibinfo {title} {Identical
  phase oscillators with global sinusoidal coupling evolve by m{\"o}bius group
  action},}\ }\href@noop {} {\bibfield  {journal} {\bibinfo  {journal} {Chaos:
  An Interdisciplinary Journal of Nonlinear Science}\ }\textbf {\bibinfo
  {volume} {19}},\ \bibinfo {pages} {043104} (\bibinfo {year}
  {2009})}\BibitemShut {NoStop}%
\bibitem [{\citenamefont {Gong}\ and\ \citenamefont
  {Pikovsky}(2019)}]{gong2019low}%
  \BibitemOpen
  \bibfield  {author} {\bibinfo {author} {\bibfnamefont {Chen~Chris}\
  \bibnamefont {Gong}}\ and\ \bibinfo {author} {\bibfnamefont {Arkady}\
  \bibnamefont {Pikovsky}},\ }\bibfield  {title} {\enquote {\bibinfo {title}
  {Low-dimensional dynamics for higher-order harmonic, globally coupled
  phase-oscillator ensembles},}\ }\href@noop {} {\bibfield  {journal} {\bibinfo
   {journal} {Physical Review E}\ }\textbf {\bibinfo {volume} {100}},\ \bibinfo
  {pages} {062210} (\bibinfo {year} {2019})}\BibitemShut {NoStop}%
\bibitem [{\citenamefont {Skardal}\ and\ \citenamefont
  {Arenas}(2019)}]{skardal2019abrupt}%
  \BibitemOpen
  \bibfield  {author} {\bibinfo {author} {\bibfnamefont {Per~Sebastian}\
  \bibnamefont {Skardal}}\ and\ \bibinfo {author} {\bibfnamefont {Alex}\
  \bibnamefont {Arenas}},\ }\bibfield  {title} {\enquote {\bibinfo {title}
  {Abrupt desynchronization and extensive multistability in globally coupled
  oscillator simplexes},}\ }\href@noop {} {\bibfield  {journal} {\bibinfo
  {journal} {Physical {R}eview {L}etters}\ }\textbf {\bibinfo {volume} {122}},\
  \bibinfo {pages} {248301} (\bibinfo {year} {2019})}\BibitemShut {NoStop}%
\bibitem [{\citenamefont {de~Arruda}\ \emph {et~al.}(2020)\citenamefont
  {de~Arruda}, \citenamefont {Petri},\ and\ \citenamefont
  {Moreno}}]{arruda2020social}%
  \BibitemOpen
  \bibfield  {author} {\bibinfo {author} {\bibfnamefont {Guilherme~Ferraz}\
  \bibnamefont {de~Arruda}}, \bibinfo {author} {\bibfnamefont {Giovanni}\
  \bibnamefont {Petri}}, \ and\ \bibinfo {author} {\bibfnamefont {Yamir}\
  \bibnamefont {Moreno}},\ }\bibfield  {title} {\enquote {\bibinfo {title}
  {Social contagion models on hypergraphs},}\ }\href@noop {} {\bibfield
  {journal} {\bibinfo  {journal} {Physical Review Research}\ }\textbf {\bibinfo
  {volume} {2}},\ \bibinfo {pages} {023032} (\bibinfo {year}
  {2020})}\BibitemShut {NoStop}%
\bibitem [{\citenamefont {Mill{\'a}n}\ \emph {et~al.}(2020)\citenamefont
  {Mill{\'a}n}, \citenamefont {Torres},\ and\ \citenamefont
  {Bianconi}}]{millan2020explosive}%
  \BibitemOpen
  \bibfield  {author} {\bibinfo {author} {\bibfnamefont {Ana~P}\ \bibnamefont
  {Mill{\'a}n}}, \bibinfo {author} {\bibfnamefont {Joaqu{\'\i}n~J}\
  \bibnamefont {Torres}}, \ and\ \bibinfo {author} {\bibfnamefont {Ginestra}\
  \bibnamefont {Bianconi}},\ }\bibfield  {title} {\enquote {\bibinfo {title}
  {Explosive higher-order {K}uramoto dynamics on simplicial complexes},}\
  }\href@noop {} {\bibfield  {journal} {\bibinfo  {journal} {Physical Review
  Letters}\ }\textbf {\bibinfo {volume} {124}},\ \bibinfo {pages} {218301}
  (\bibinfo {year} {2020})}\BibitemShut {NoStop}%
\bibitem [{\citenamefont {Totz}\ \emph {et~al.}(2018)\citenamefont {Totz},
  \citenamefont {Rode}, \citenamefont {Tinsley}, \citenamefont {Showalter},\
  and\ \citenamefont {Engel}}]{totz2018spiral}%
  \BibitemOpen
  \bibfield  {author} {\bibinfo {author} {\bibfnamefont {Jan~Frederik}\
  \bibnamefont {Totz}}, \bibinfo {author} {\bibfnamefont {Julian}\ \bibnamefont
  {Rode}}, \bibinfo {author} {\bibfnamefont {Mark~R}\ \bibnamefont {Tinsley}},
  \bibinfo {author} {\bibfnamefont {Kenneth}\ \bibnamefont {Showalter}}, \ and\
  \bibinfo {author} {\bibfnamefont {Harald}\ \bibnamefont {Engel}},\ }\bibfield
   {title} {\enquote {\bibinfo {title} {Spiral wave chimera states in large
  populations of coupled chemical oscillators},}\ }\href@noop {} {\bibfield
  {journal} {\bibinfo  {journal} {Nature Physics}\ }\textbf {\bibinfo {volume}
  {14}},\ \bibinfo {pages} {282--285} (\bibinfo {year} {2018})}\BibitemShut
  {NoStop}%
\bibitem [{\citenamefont {C{\u{a}}lug{\u{a}}ru}\ \emph
  {et~al.}(2020)\citenamefont {C{\u{a}}lug{\u{a}}ru}, \citenamefont {Totz},
  \citenamefont {Martens},\ and\ \citenamefont {Engel}}]{cualuguaru2020first}%
  \BibitemOpen
  \bibfield  {author} {\bibinfo {author} {\bibfnamefont {Dumitru}\ \bibnamefont
  {C{\u{a}}lug{\u{a}}ru}}, \bibinfo {author} {\bibfnamefont {Jan~Frederik}\
  \bibnamefont {Totz}}, \bibinfo {author} {\bibfnamefont {Erik~A}\ \bibnamefont
  {Martens}}, \ and\ \bibinfo {author} {\bibfnamefont {Harald}\ \bibnamefont
  {Engel}},\ }\bibfield  {title} {\enquote {\bibinfo {title} {First-order
  synchronization transition in a large population of strongly coupled
  relaxation oscillators},}\ }\href@noop {} {\bibfield  {journal} {\bibinfo
  {journal} {Science advances}\ }\textbf {\bibinfo {volume} {6}},\ \bibinfo
  {pages} {eabb2637} (\bibinfo {year} {2020})}\BibitemShut {NoStop}%
\bibitem [{\citenamefont {Hagerstrom}\ \emph {et~al.}(2012)\citenamefont
  {Hagerstrom}, \citenamefont {Murphy}, \citenamefont {Roy}, \citenamefont
  {H{\"o}vel}, \citenamefont {Omelchenko},\ and\ \citenamefont
  {Sch{\"o}ll}}]{hagerstrom2012experimental}%
  \BibitemOpen
  \bibfield  {author} {\bibinfo {author} {\bibfnamefont {Aaron~M}\ \bibnamefont
  {Hagerstrom}}, \bibinfo {author} {\bibfnamefont {Thomas~E}\ \bibnamefont
  {Murphy}}, \bibinfo {author} {\bibfnamefont {Rajarshi}\ \bibnamefont {Roy}},
  \bibinfo {author} {\bibfnamefont {Philipp}\ \bibnamefont {H{\"o}vel}},
  \bibinfo {author} {\bibfnamefont {Iryna}\ \bibnamefont {Omelchenko}}, \ and\
  \bibinfo {author} {\bibfnamefont {Eckehard}\ \bibnamefont {Sch{\"o}ll}},\
  }\bibfield  {title} {\enquote {\bibinfo {title} {Experimental observation of
  chimeras in coupled-map lattices},}\ }\href@noop {} {\bibfield  {journal}
  {\bibinfo  {journal} {Nature Physics}\ }\textbf {\bibinfo {volume} {8}},\
  \bibinfo {pages} {658--661} (\bibinfo {year} {2012})}\BibitemShut {NoStop}%
\end{thebibliography}%
%\end{thebibliography}
%\bibliographystyle{apsrev4-1}

\end{document}